\documentclass[twocolumn,times]{aastex7}
\usepackage{amsmath,amssymb,mathtools,float,multirow}
\usepackage[T1]{fontenc}

\definecolor{plotorange}{HTML}{DA674D}
\definecolor{plotblue}{HTML}{1AAD85}

\newcommand{\xspec}{\texttt{Xspec}}

\newcommand{\rin}[1][]{%
  \ifx\relax#1\relax
    \ensuremath{r_{\rm in}}%
  \else
    \ensuremath{r_{{\rm in}, #1}}%
  \fi
}

\newcommand{\rout}[1][]{%
  \ifx\relax#1\relax
    \ensuremath{r_{\rm out}}%
  \else
    \ensuremath{r_{{\rm out}, #1}}%
  \fi
}

\newcommand{\rg}[1][]{%
  \ifx\relax#1\relax
    \ensuremath{r_{g}}%
  \else
    \ensuremath{r_{g, #1}}%
  \fi
}

\newcommand{\risco}[1][]{%
  \ifx\relax#1\relax
    \ensuremath{r_{\rm ISCO}}%
  \else
    \ensuremath{r_{#1{\rm (ISCO)}}}%
  \fi
}

\newcommand{\rl}[1][]{%
  \ifx\relax#1\relax
    \ensuremath{r_{\rm L}}%
  \else
    \ensuremath{r_{{\rm L}, #1}}%
  \fi
}

\newcommand{\rms}[1][]{%
  \ifx\relax#1\relax
    \ensuremath{r_{\rm MS}}%
  \else
    \ensuremath{r_{#1{\rm (MS)}}}%
  \fi
}

\newcommand{\ltot}{\ensuremath{\lambda_{\rm tot}}}

\newcommand{\logxi}[1][]{%
  \ifx\relax#1\relax
    \ensuremath{{\rm log} (\xi)}%
  \else
    \ensuremath{{\rm log} (\xi_{#1})}%
  \fi
}

\newcommand{\Afe}[1][]{%
  \ifx\relax#1\relax
    \ensuremath{A_{\rm Fe}}%
  \else
    \ensuremath{[A_{\rm Fe}]_{#1}}%
  \fi
}

\newcommand{\degree}{\ensuremath{^{\circ}}}

\newcommand{\lisa}{\textit{LISA}}
\newcommand{\strobex}{\textit{STROBE-X}}
\newcommand{\athena}{\textit{NewAthena}}
\newcommand{\axis}{\textit{AXIS}}
\newcommand{\hexp}{\textit{HEX-P}}

\newcommand{\fe}{Fe~K$\alpha$}

\newcommand{\add}[1]{{#1}}

\begin{document}

\title{X-Ray Reflection Signatures of Supermassive Black Hole Binaries}



\author[0009-0004-2625-5527]{Julie Malewicz}
\affiliation{Center for Relativistic Astrophysics, School of Physics,
  Georgia Institute of Technology, 837 State Street NW, Atlanta, GA 30332-0430, USA}
\email[show]{jlm@gatech.edu}

\author[0000-0001-8128-6976]{David R.\  Ballantyne}
\affiliation{Center for Relativistic Astrophysics, School of Physics,
  Georgia Institute of Technology, 837 State Street NW, Atlanta, GA 30332-0430, USA}
\email{david.ballantyne@gatech.edu}

\author[0000-0002-7835-7814]{Tamara Bogdanovi\'c}
\affiliation{Center for Relativistic Astrophysics, School of Physics,
  Georgia Institute of Technology, 837 State Street NW, Atlanta, GA 30332-0430, USA}
\email{tamarab@gatech.edu}

\author[0000-0003-2663-1954]{Laura Brenneman}
\affiliation{Center for Astrophysics, Harvard-Smithsonian, 60 Garden Street, Cambridge, MA 02138, USA}
\email{lbrenneman@cfa.harvard.edu}

\author[0000-0003-4583-9048]{Thomas Dauser}
\affiliation{Dr. Karl Remeis-Observatory \& ECAP, FAU Erlangen-Nürnberg,
  Sternwartstr. 7, 96049 Bamberg, Germany}
\email{thomas.dauser@sternwarte.uni-erlangen.de}

%
%

\begin{abstract}
We investigate the presence of supermassive black hole (SMBH) binary signatures and the feasibility of identifying them through X-ray reflection spectra. The X-ray emitting region is modeled as a set of two mini-disks bound to the individual SMBHs separated by $100\,GM/c^2$ and the spectra calculated as a function of the mass, mass ratio, and total accretion rate of the binary. The X-ray reflection features are strongly influenced by the accretion-inversion phenomenon expected in SMBH binaries, which results in a wide range of ionization conditions in the two mini-disks. These are imprinted in the resulting composite spectra and the double-peaked and time-variable relativistic \fe\ line profiles. To test whether these features can be used as evidence for the presence of an SMBH binary, we fit mock 100\,ks observations with a single AGN model. For a $10^9\,M_\odot$ binary targeted by Pulsar Timing Arrays (PTAs), at $z=0.1$ the single AGN model clearly fails to fit the data, while at $z=1$ the fit is acceptable but unable to converge on the SMBH spin. For a $10^6\,M_\odot$ binary, a progenitor of a \textit{Laser Interferometer Space Antenna} ({\it LISA}) source, spectral fitting is only possible at $z=0.1$, with the outcomes similar to the PTA binary at $z=1$. We also find that PTA binaries can be expected to show a distinct X-ray spectral variability in multi-epoch observations, whereas for \lisa\ precursors, orbital averaging results in the loss of spectral variability signatures.

\end{abstract}



%
%

\section{Introduction}\label{sec:intro}

Supermassive black hole (SMBH) binaries are expected to be the natural
byproducts of standard hierarchical galaxy formation
\citep[e.g.,][]{begelman1980,dimatteo2005,hopkins2008}. A bound binary
will form after a galaxy merger when the two SMBHs reach a separation
$\lesssim 1$~pc, after which the orbit slowly tightens through three-body interactions
with nearby stars and local gas flows
\citep[e.g.,][]{antonini2015}. Finally, at separations of $\sim
10^{-3}$~pc, the emission of gravitational waves (GWs) begins to dominate the orbital
evolution \citep[e.g.,][]{peters1963,mangiagli2022}, driving the
binary towards merger. Detecting electromagnetic (EM) signatures
of SMBH binaries during their orbital decay is of particular interest
in the era of multimessenger astronomy, as they will provide crucial
information and context for interpreting GW signatures
from SMBH mergers \citep[e.g.,][]{bogdanovic2022}. Indeed, the recent
detection of a stochastic GW background by millisecond Pulsar Timing
Arrays \citep[PTAs;][]{agazie2023,antoniadis2023,reardon2023,xu2023} could be explained by
numerous SMBH binaries locked in tight, sub-parsec orbits
\citep[e.g.,][]{agazie2023b}. Searches for the EM signatures of this population of SMBH
binaries is crucial to understanding this background. The EM identification of SMBH binaries with sub-parsec separations must rely on indirect methods, as they cannot be spatially resolved using imaging with current instruments.

Galaxy mergers not only produce SMBH binaries, but also drive a significant
inflow of gas to the central regions of the merger remnant
\citep[e.g.,][]{barnes1991, dimatteo2005, hopkins2005, kazantzidis2005}. This provides a favorable environment for the two SMBHs to
accrete efficiently and shine as active galactic nuclei (AGNs) during their
evolution
\citep[e.g.,][]{volonteri2003,sesana2012,derosa2019}. AGN activity
often produces emission lines from ionized gas bound to the SMBH;
therefore, searches for changes or unusual offsets in these emission
line profiles may be an avenue to find evidence of a SMBH binary
\citep[e.g.,][]{bogdanovic2009, eracleous2012,runnoe2015}. While on the order of one hundred SMBH binary candidates have
been identified using the profiles of broad optical emission lines,
uncertainties in the dynamics of broad-line region clouds, and the
challenges of separating the signatures of a potential binary from a
recoiling single AGN, make the interpretation of the results
ambiguous \citep[e.g.,][]{husemann20}.

AGNs also often exhibit a fluorescent \fe\ emission from the inner accretion
disk in their X-ray spectra
\citep[e.g.,][]{tanaka1995,brenneman2006,reynolds2013}. This broad line
originates from gas tightly orbiting the SMBH, sculpting its profile
through relativistic effects \citep[e.g.,][]{fabian1989}. The \fe\ line is
one part of an overall reflection spectrum emitted by the accretion
disk due to its irradiation by a central, hot corona
\citep[e.g.,][]{garcia2010,dauser2014,ballantyne2017}. Recent
simulations of SMBH binaries show that each SMBH will be accreting
from a `mini-disk' that is fed from a larger circumbinary disk
\citep[e.g.,][]{farris2014,bowen2017,tang2018,avara2024}. If the mini-disks produce X-ray emitting coronae similar to those of single
AGNs, then each accretion disk of the binary will produce its own
reflection spectrum, including a relativistically broadened \fe\ line. The composite reflection spectrum may therefore contain signatures of the binary
that could be identified with a sufficiently high-quality X-ray
observation. Moreover, since the shape of the \fe\ line contains
information on the spin of the central SMBH
\citep[e.g.,][]{reynolds2008}, fundamental properties of both SMBHs could, in principle, be accessible through fitting the combined
\fe\ line.

\add{Efforts to identify binary candidates in the X-ray regime have not yet found any definitive cases, underlining the fact that clear observational evidence for binarity may be challenging to detect with current instrumentation. For instance, while it is generally expected that the light originating from binary mini-disks may exhibit periodic variations in luminosity, a systematic search for such signatures in the X-ray light curves of hundreds of AGN in the Swift-BAT survey found no conclusive evidence for periodicity \citep{liu2020}. Similarly, an unambiguous example of a double-peaked \fe\ line has yet to be confirmed (e.g., MCG+11-11-032; \citealt{severgnini2018,foord2025}).} 
Previous \add{theoretical models for} the X-ray spectra in SMBH binaries have either focused on the modeling of the \fe\ line in isolation, without accounting for the different ionization properties expected from mini-disks in unequal mass binaries \citep{yu2001,popovic2012,jovanovic2014,jovanovic2020}, or treated the emission as thermal in origin, without the presence of the full reflection spectrum \citep[e.g.,][]{farris2015,dascoli2018}.  
%
Here, we study the composite, relativistic, ionized X-ray reflection
spectra, from roughly 0.1 to 100~keV, expected from two
gravitationally bound SMBH at a separation of $100$ gravitational
radii, taking into account the expected size of the mini-disks. We
show how the composite spectra change as a function of the accretion
rate onto the binary and the mass ratio of the two SMBHs. Finally, we
 evaluate the prospects for detection of any binary signatures by
 considering mock observations from the next-generation X-ray missions
 currently in development.

 In the next section, we lay out the details of our methods to model
 an accreting SMBH binary and compute its composite X-ray
 spectrum. Section~\ref{sec:properties} presents the resulting
 composite spectra for various combinations of mass ratio and
 total accretion rate onto the binary. Section~\ref{sec:prospects} describes the simulated
 observations of a fiducial SMBH binary by 4 well-studied X-ray telescope
 concepts (\athena, \axis, \hexp\ and \strobex), as
 well as a preliminary time variability study of a promising
 configuration with \athena. A discussion of the implications and
 limitations of our results is found in Sect.~\ref{sec:discussion},
 and overall conclusions are presented in Sect.~\ref{sec:concl}.

%
%

\section{Methods}
\label{sec:methods}

\subsection{Composition of Binary System}
\label{sub:composition}
We model SMBH binary systems composed of two near-maximally spinning\footnote{\add{This choice is motivated by the high fraction of high spins measured in bright AGN \citep[e.g.,][]{reynolds2021}, and is discussed in Section \ref{sub:disc_binary}.}}
(i.e., spin parameter $a=0.99$) SMBHs with mass ratio $q=m_2/m_1$
where $0.1 \leq q \leq 0.9$, in a circular orbit separated by a distance of $s=100\,r_g$
($r_g=GM/c^2$ is the gravitational radius for a binary with total mass
$M=m_1+m_2$). The total mass of the binary is set to either
$10^9$~M$_{\odot}$ or $10^6$~M$_{\odot}$. The $M=10^9$~M$_\odot$
binaries at $s=100\,r_g$ emit GWs within the PTA band \citep[e.g.,][]{sesana2012}, while
$M=10^6$~M$_\odot$ binaries at this separation should enter the
\lisa\ sensitivity band by the time they merge. Following \citet{peters1964}, the characteristic orbital and GW merger time scale for such binaries is
\begin{equation}
    t_{\rm orb} =  0.36\,{\rm days} \left(\frac{M}{10^6\,M_\odot} \right)
    \left(\frac{s}{100\,r_{\rm g}}\right)^{3/2}\;\;,
    \label{eq:torbital}
\end{equation}
and 
\begin{equation}
    t_{\rm gw} = 0.31\,{\rm yr}\, \frac{(1+q)^2}{q} 
    \left(\frac{M}{10^6\,M_\odot} \right)
    \left(\frac{s}{100\,r_{\rm g}}\right)^4 \;\;,
    \label{eq_tgw}
\end{equation}
The difference in those two time scales for the SMBH binaries at this characteristic separation allows us to consider $s$ as time-independent in our calculations. In
what follows, we refer to the $10^9$~M$_{\odot}$ binaries as `PTA'
binaries and the $10^6$~M$_{\odot}$ binaries as `\lisa' binaries.

Each SMBH in the binary is accreting through a mini-disk, and both
objects are orbiting within a low-density binary cavity surrounded by
a circumbinary disk (CBD). The radius of the cavity from the binary
center of mass to the inner edge of the CBD is equal to 2$s$ \citep[e.g.,][]{macfayden2008,roedig2011}.

We assume both mini-disks are fixed to
their central SMBHs and are co-aligned with the binary orbital plane,
extending radially from \rin$=1.5$~$r_g$ (the
innermost stable circular orbit; ISCO) to \rout, where both mini-disks are
tidally truncated due to orbital resonances in the binary orbital
cavity
\citep{papaloizou1977,lin1979,paczynski1980,roedig2014}. As
opposed to the accretion disk of an isolated SMBH, SMBH binaries
evolve in a time-dependent gravitational potential that truncates
the attached mini-disks by destabilizing particle orbits in the binary
cavity. Following \citet{pichardo2005},
the outermost stable particle orbit of the SMBHs in a circular binary
are
\begin{equation}
    \rout \approx 0.733 \left( \frac{q}{1+q} \right)^{0.07} \rl,
\label{eq:rout}
\end{equation}
where \rl\ is the Roche lobe radius of each SMBH \citep{eggleton1983}. For example, $\rout \approx 0.62\,\rl - 0.70\,\rl$ for a range of mass ratios between $q=0.1$ and 1, respectively. At the same time, the Roche lobe radius of the primary (secondary) SMBH varies between $0.58s - 0.38s$ ($0.21s-0.38s$), respectively.

\subsection{Accretion Rates}
\label{sub:accretion}
Gas supplied by the CBD flows at a steady rate $\dot{M}$ onto the mini-disks'
outer edges. The accretion rates onto the primary and secondary SMBHs
of the binary, $\dot{m}_1$ and $\dot{m}_2$ (where $\dot{M}=\dot{m}_1+\dot{m}_2$), depend on the mass ratio
$q$ with the larger accretion rate corresponding to the lighter SMBH. This effect, known as accretion inversion
\citep{artymowicz1994, guenther2002, hayasaki2006, roedig2011,
  farris2015}, occurs because the lighter SMBH orbits closer to the
inner rim of the CBD. 
\citet{kelley2019} describe this effect by fitting the simulation data
of \citet{farris2015}, finding
\begin{equation}
  \label{eq:mdot}
\dot{m}_2 = \dot{m}_1 \left(\frac{50}{(12 q)^{3.5} + (12 q)^{-3.5}} \right).
\end{equation}
We adopt this description in our calculation and further assume that $\dot{m}_1$ and $\dot{m}_2$ are constant during the binary orbit, for simplicity.

The total accretion rate onto the binary is characterized by its
Eddington ratio,
\begin{equation}
  \label{eq:lambdatot}
  \lambda_{\mathrm{tot}} = {\dot{M} \over \dot{M}_{\mathrm{Edd,tot}}},
\end{equation}
where $\dot{M}_{\mathrm{Edd,tot}}=(2.2\times 10^{-8}$~M$_{\odot} \; $~yr$^{-1} ) (0.1/\varepsilon_{\rm rad}) \; (M/$M$_{\odot}$) and the radiative
efficiency\footnote{While the maximum possible radiative efficiency for a SMBH with spin $a=0.99$ is $\approx 0.4$ \citep[e.g.,][]{blandford1977}, we adopt a value of $0.1$, consistent with observations of AGNs \citep[e.g.,][]{davis2011}. A larger value of $\varepsilon_{\rm rad}$ would result in rescaling of both mini-disk ionization parameters by the same factor and consequently, we do not expect it to have a significant impact on our results.} $\varepsilon_{\rm rad}$ is $0.1$. The Eddington ratios of the primary and secondary
are therefore $\lambda_1=\dot{m}_1/\dot{M}_{\mathrm{Edd,1}}$ and
$\lambda_2=\dot{m}_2/\dot{M}_{\mathrm{Edd,2}}$. Defining
$\kappa=\dot{m}_2/\dot{m}_1$, expressions for $\lambda_1$ and
$\lambda_2$ can be derived in terms of $\kappa$, $q$ and
\ltot:
\begin{equation}
  \label{eq:lambda1}
  \lambda_1 = {1 + q \over 1 + \kappa} \lambda_{\mathrm{tot}}
\end{equation}
and
\begin{equation}
  \label{eq:lambda2}
  \lambda_2 = {1 +q \over q} {\kappa \over 1+\kappa}
  \lambda_{\mathrm{tot}}.
\end{equation}
As $\kappa$ is only a function of $q$, the Eddington ratios onto both
SMBHs are determined by specifying $q$ and \ltot, and are independent
of the total mass of the binary $M$.

\begin{figure*}[t!]
    \centering
    \includegraphics[width=0.9\linewidth]{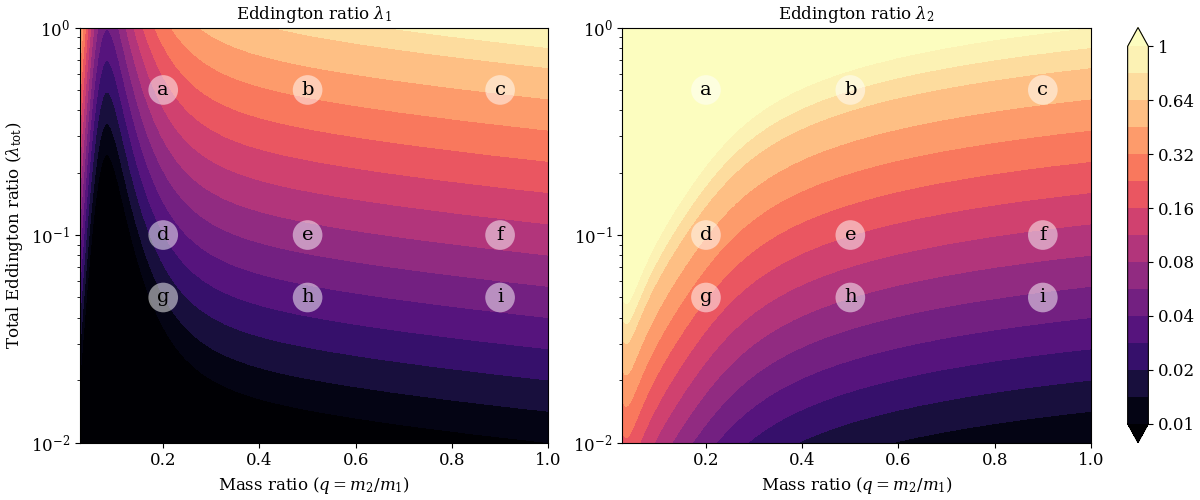}
    \caption{The color map shows the individual Eddington ratios for
      the primary (left) and secondary (right) SMBHs as a function of
      binary mass ratio ($q=m_2/m_1$) and the total Eddington ratio
      onto the binary (\ltot; Eq. \ref{eq:lambdatot}). The accretion
      inversion phenomenon is easily seen in these plots, in particular at $q
      \lesssim 0.1$ where $\lambda_1 < 0.01$ and $\lambda_2$ is super-Eddington even if $\lambda_{\mathrm{tot}} \sim
    0.1$. The letters correspond to specific model parameters discussed in Sect.~\ref{sec:properties} and
    Table~\ref{tab:param}.}
    \label{fig:eddington_contour}
\end{figure*}

Color maps of $\lambda_1$ and $\lambda_2$ in the
$(q,\lambda_{\mathrm{tot}})$ plane are shown in
Figure~\ref{fig:eddington_contour}.
%
The effects of accretion inversion are clearly seen in the figures,
and are most acute when $q \la 0.1$ and the secondary is orbiting
close to the inner edge of the CBD. In this scenario, $\lambda_2$ can
be driven into the super-Eddington regime even if
$\lambda_{\mathrm{tot}} \sim 0.1$. The strong variation of the 
Eddington ratios across the parameter space impacts the likelihood of observing a significant X-ray reflection signal from a potential binary and results in a wide range of ionization conditions. X-ray reflection from the inner accretion disk requires the
presence of Compton-thick gas. When Eddington ratios get too low ($\la
0.01$) or too high ($\ga 1$), accretion disks are expected to become
geometrically thick and thus less dense \citep[e.g.,][]{blaes14}, and less
susceptible to producing an X-ray reflection spectrum. We
therefore focus (with one exception) on $(q,\lambda_{\mathrm{tot}})$ values
where the Eddington ratios for both SMBHs are between $0.01$ and $1$
and a geometrically-thin, optically-thick accretion disk is likely to
exist around each object. The specific values chosen are indicated by
the letters in Fig.~\ref{fig:eddington_contour} and are shown in
Table~\ref{tab:param}.
\begin{deluxetable}{c|cc||cc|cc}[t!]
\tabletypesize{}
\tablewidth{0pt} 
\tablecaption{Eddington ratios and peak disk ionization parameters for
  each SMBH binary configuration considered in
  Sect.~\ref{sec:properties}. The letters correspond to the locations
  marked in Figs.~\ref{fig:eddington_contour}
  and~\ref{fig:ionization_contour}. \label{tab:param}}
\tablehead{
\colhead{} & \colhead{\ltot} & \colhead{$q$} & \colhead{$\lambda_1$}& \colhead{\logxi$_1$} & \colhead{$\lambda_2$} & \colhead{\logxi$_2$}
} 
\startdata
(a) & 0.5  & 2/10 & 0.14 & 2.90 & 2.30 & 6.01$^{\dag}$ \\
(b) & 0.5  & 5/10 & 0.35 & 3.94 & 0.80 & 4.87$^{\dag}$ \\
(c) & 0.5  & 9/10 & 0.47 & 4.28 & 0.53 & 4.41 \\
(d) & 0.1  & 2/10 & 0.03 & 0.98 & 0.46 & 4.26 \\
(e) & 0.1  & 5/10 & 0.07 & 2.09 & 0.16 & 3.07 \\
(f) & 0.1  & 9/10 & 0.09 & 2.45 & 0.11 & 2.59 \\
(g) & 0.05 & 2/10 & 0.01 & 0.13 & 0.23 & 3.48 \\
(h) & 0.05 & 5/10 & 0.04 & 1.26 & 0.08 & 2.26 \\
(i) & 0.05 & 9/10 & 0.05 & 1.62 & 0.05 & 1.77 \\[5pt]
\enddata
\tablecomments{$^\dag$ Replaced with $4.70$ for calculation with the
  \texttt{relxilllpCp} model.} 
\end{deluxetable}
Configuration (a) does have $\lambda_2=2.3$, but as this is only
slightly super-Eddington we include it in our sample in order to
consider uniform coverage of the $(q,\lambda_{\mathrm{tot}})$ plane.

\subsection{X-ray Reflection Calculation}
\label{sub:reflection_calculation}

The accretion disk around each SMBH in the binary is assumed to be irradiated by
a static, hard X-ray emitting corona located $10$~$r_g$ above the black
hole (i.e., a `lamppost' corona; e.g.,
\citealt{martocchia1996,dauser2013}). We neglect any X-ray flux from
the corona of the primary that may strike the disk of the secondary
(and vice versa). The relativistic ionized X-ray reflection spectrum from each
accretion disk is computed using the \texttt{relxilllpCp} model
\citep[e.g.,][]{dauser2016} within \xspec\ \citep{xspec}. As described
in Sect.~\ref{sub:composition}, the SMBH spins are flixed at $a=0.99$
and the disks extend from \rin\ to $r_{\mathrm{out}}$
(Eq.~\ref{eq:rout}). Each corona has an electron temperature of
$60$~keV and produces a Comptonization spectrum
with photon-index $\Gamma=2$, consistent with values commonly observed
from single AGNs \citep[e.g.,][]{fabian2015}. 
We assume a disk inclination of 30 degrees (we discuss the emission geometry in more detail in Section~\ref{sub:geometry}) and adopt Solar abundances.

The strength of the spectral features in reflection spectra, including
the \fe\ line, are strong functions of the ionization state of the gas
at the surface of the accretion disk
\citep[e.g.,][]{ross2005,garcia2010}. If the disk surface is strongly
illuminated and becomes highly ionized, then it becomes
effectively a mirror, reflecting the irradiating spectrum outwards,
augmented with only very weak and highly Comptonized spectral
features. The properties of a reflection spectrum can be described
using the ionization parameter, $\xi = 4 \pi F_X/n_H$, where $F_X$ is
the X-ray flux irradiating gas with hydrogen number density
$n_H$. Gas with $\log \xi \ga 4$ is considered highly ionized and
yields very weak reflection features \citep[e.g.,][]{garcia2010}.

A central compact corona, such as the lamppost model, will produce a
radially dependent $F_X(r)$ \citep[e.g.,][]{fukumura2007,dauser2013}
which, when combined with the density profile $n_H(r)$, will
generate an ionization gradient across the disk
\citep[e.g.,][]{ballantyne2017}. The total reflection spectrum from
each mini-disk is found by integrating the reflection spectra over
disk radii, taking into account the relativistic shifts that blur the
spectra at each radius \citep[e.g.,][]{fabian1989}. As the inner disk
is more strongly illuminated than the outer disk, the ionization
parameter at the inner edge is crucial in determining the
properties of the total reflection spectrum. 

The \texttt{relxilllpCp} model includes the effects of this
ionization gradient (assuming a \citealt{shakurasunyaev1973} radial density profile) when calculating the total relativistic
reflection spectrum of each mini-disk. The density at the inner edge
of each disk is set to $10^{15}$~cm$^{-3}$ (the impact of this
assumption is discussed in Sect.~\ref{sub:limitations}). The peak
ionization parameter at a distance $(11/9)^2$~\rin, denoted $\xi_1$ and
$\xi_2$, must also be specified for each mini-disk. These values are
computed using the analytical ionization gradient formula for a
lamppost corona derived by \citet{ballantyne2017}. This formula
depends on the Eddington ratio of each mini-disk, the radiative
efficiency (set to $0.1$, as before), the Shakura-Sunyaev
$\alpha$-parameter (set to $0.1$), and is
independent of SMBH mass. The \citet{ballantyne2017} equation also
makes use of the coronal dissipation fraction $f_X$ \citep{svensson1994}
to convert from the bolometric luminosity, powered by accretion, to the X-ray flux. Here,
we connect $f_X$ to the X-ray bolometric correction, $k_{X}$, via
$f_X=1/k_X$, and use the \citet{duras2020} fitting formula to calculate
$k_X$ at the Eddington ratio of each mini-disk. 

\begin{figure*}[t!]
  \centering
  \includegraphics[width=0.9\linewidth]{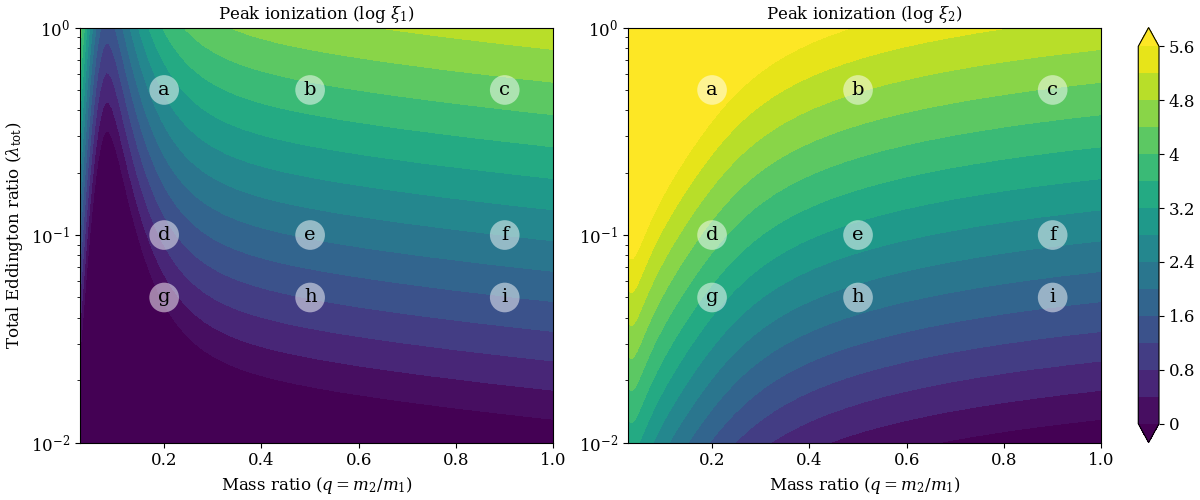}
  \caption{The color maps show the peak ionization parameter in the
    mini-disks around the primary ($\log \xi_1$; left panel) and
    secondary ($\log \xi_2$; right panel) SMBHs in the $(q,
    \lambda_{\mathrm{tot}})$ plane. The ionization parameters are
    calculated using the \citet{ballantyne2017} ionization gradient
    formula evaluated at a radius of $(11/9)^2$~\rin, which are then
    input to
    \texttt{relxilllpCp} to compute the relativistic ionized
    reflection spectrum emitted by each mini-disk. The accretion inversion phenomenon,
    combined with the $\lambda^3$ dependence of the
    \citet{ballantyne2017} equation, causes a large variation of
    $\xi_{1,2}$ through the parameter space with the mini-disk of the
    secondary expected to be more highly ionized than the primary. The values of $\log
    \xi_1$ and $\log \xi_2$ at the location of the letters in the
    plots are listed in Table~\ref{tab:param}.}
  \label{fig:ionization_contour}
\end{figure*}

The resulting values of $\log \xi_1$ and $\log \xi_2$ computed using the
\citet{ballantyne2017} formula are shown as
functions of $q$ and \ltot\ in Figure~\ref{fig:ionization_contour}. 
%
The color maps show the impact of the accretion inversion effect on
the ionization parameters, which is exacerbated by the strong
dependence of $\xi$ on $\lambda$ predicted by the
\citet{ballantyne2017} equation ($\xi \propto \lambda^3$). As a
result, the ionization state of the mini-disks will change
by orders of magnitude in response to only moderate differences in $q$
or \ltot. This will lead to a wide range of possible \fe\ line
strengths and shapes, as well as
other reflection signatures, in the composite spectra of SMBH
binaries, especially when $q \la 0.4$. The accretion inversion also
causes the mini-disk of the secondary SMBH to be more highly ionized
than the primary mini-disk. The letters on the plots show
the 9 $(q,\lambda_{\mathrm{tot}})$ locations from
Fig.~\ref{fig:eddington_contour} and the associated $\log \xi_1$
and $\log \xi_2$ values are found in Table~\ref{tab:param}. The
peak ionization parameters span the range from $\log \xi =0.05$ to
$>5$; however, the maximum $\log \xi$ allowed by \texttt{relxilllpCp}
is $4.70$, so any peak ionization parameter $> 10^{4.7}$ is replaced
with this value. In practice, once the illuminated gas is fully
ionized (at $\log \xi \ga 4$; \citealt{garcia2010}), there is little
difference in the predicted reflection spectra for any higher value of
$\xi$. Therefore, the use of the $\log \xi = 4.7$ upper-limit will not
affect the properties of the predicted SMBH binary spectra.

Now armed with the values of $\log \xi_1$ and $\log \xi_2$,
\texttt{relxilllpCp} models can be computed for each mini-disk in the
binaries listed in Table~\ref{tab:param}. In addition to the
parameters already described, we set the redshifts to zero, allow for
the effects of returning radiation \citep{dauser2022}, and include the
correct contribution from the continuum in a lamppost geometry (i.e.,
\texttt{refl\_frac=1} and \texttt{switch\_reflfrac\_boost=1}). For plotting purposes,
all spectra are computed from $0.02$ --$200$~keV in 2000 logarithmically
spaced bins.

\subsection{Emission Geometry and Calculation of the 
  Composite Spectra}
\label{sub:geometry}

At this stage, we have calculated the relativistic ionized reflection
spectrum from the mini-disks of 9 SMBH binary configurations described by the
$(q,\lambda_{\mathrm{tot}})$ parameters in
Table~\ref{tab:param}. However, each SMBH in a binary will be
moving relative to each other, leading to velocity shifts that must be
taken into account when combining the two X-ray spectra to form the
composite. We therefore must specify details of the geometry
of the SMBH binaries in order to account for these velocity shifts.

\begin{figure}[t!]
    \centering
    \includegraphics[width=1\linewidth]{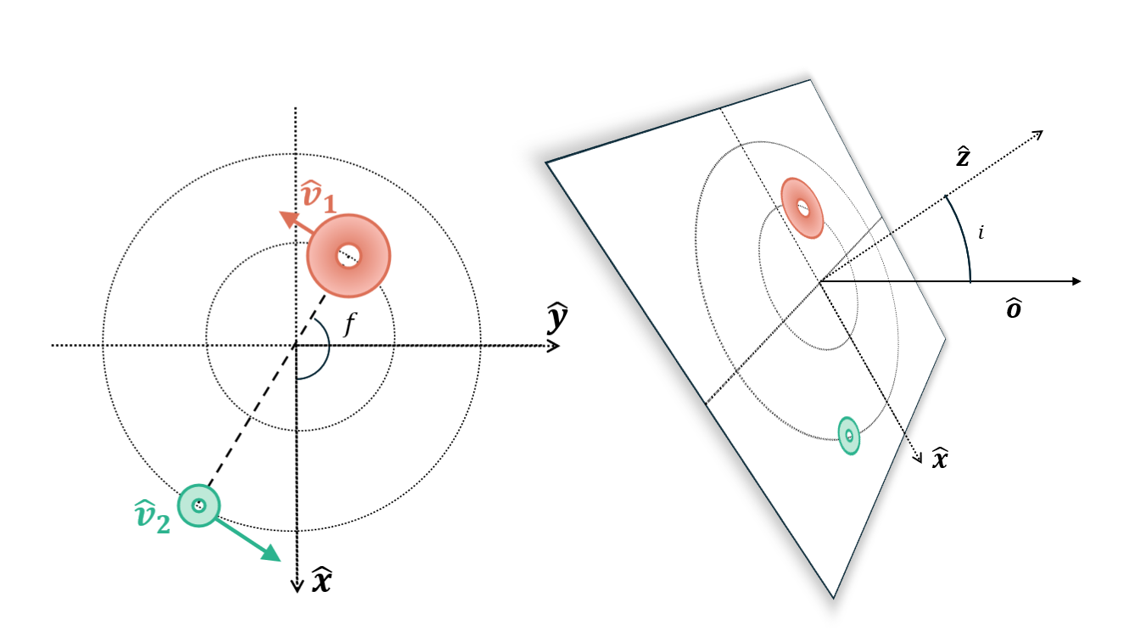}
    \caption{Illustration of the binary emission geometry. The SMBHs are on
      concentric circular orbits on the $xy$-plane, where $\hat{x}$ is
      defined as the projection of the $\hat{o}$ vector (from binary
      center of mass to the observer) onto the binary orbital
      plane. $\hat{z}$ is defined as the angular momentum vector of the
      binary, perpendicular to the plane and forming an angle $i$ with
      the observer vector. The orbital phase $f$ is defined as the
      angle between the $\hat{x}$ axis and the primary black hole. 
    }
    \label{fig:geometry}
\end{figure}

Our chosen geometry is depicted in Figure \ref{fig:geometry}. Both
mini-disks are coplanar with the binary orbital plane.
%
The inclination $i$ is defined as the angle between the line of
sight $\mathbf{\hat{o}}$ (defined as originating from the binary
center of mass and pointing to the observer) and the binary
orbital angular momentum vector $\mathbf{\hat{z}}$ (normal to the
binary orbital plane). If $i \in [-90 \degree,90 \degree]$, then
$\mathbf{\hat{o}} \cdot \mathbf{\hat{z}}>0$, and the orbit appears
counterclockwise, and vice-versa.  The phase angle $f$ takes on values from 0 to $360\degree$. Conjunction,  when both orbital velocities are completely perpendicular to $\mathbf{\hat{o}}$, happens at $f=0 \degree$ (defined as the point at which the primary black hole crosses the $\mathbf{+\hat{x}}$-axis) and $f=180\degree$, when the secondary is the one crossing the $\mathbf{+\hat{x}}$ axis. In both cases, the orbital Doppler effect is minimal (but non-zero, as we take into account a transverse Doppler effect). Conversely, the opposition is defined as the two points when the BH velocities projected onto the line of sight are at their maximum. The secondary moves towards the observer at $f=90\degree$, and away from it it at $f=270\degree$. The opposite is true for the primary. The more ``edge-on'' the binary is ($i$ close to $90 \degree$), the more pronounced the Doppler shift on both spectra is expected to be. As with the \texttt{relxilllpCp} models described in Section~\ref{sub:reflection_calculation}, we adopt the inclination angle $i=30^\circ$. We defer an investigation of a wider range of inclinations to a future study and note that the adopted value affects both the relativistic blurring of spectral features and the apparent velocity offset of the two SMBHs.

The orbital speed of each of the two SMBHs relative to the binary center of mass is 
\begin{equation}
  \label{eq:v1}
  v_1 = {q \over 1 + q} \; v \;\;{\rm and}\;\; v_2 = {1 \over 1 + q} \; v,
\end{equation}
%
%
%
where $v = \sqrt{GM /s}$ is the relative orbital speed of the two SMBHs in a circular Keplerian orbit, which is equal to 0.1$c$ for $s=100~\rg$. 
The Doppler factor depends on the fraction of the orbital velocity that is parallel to the line of sight, ${\vec{v}}_{1,2} \cdot \mathbf{\hat{o}} = \pm |\vec{v}_{1,2}| \sin{i} \sin{f}$: 
\begin{equation}
  \label{eq:doppler}
  1+ z_\mathrm{orb} = \frac{\sqrt{1-(v_{1,2}/c)^2}}{1 \pm (v_{1,2}/c)\sin{i}\sin{f}}
\end{equation}
where, because we have defined $f$ around the location of the secondary \add{SM}BH, we take the negative value when computing the primary's Doppler shift. These orbital redshifts are implemented using the \texttt{zashift} function in \xspec.

%
\begin{figure*}[t!]
    \centering
    \includegraphics[trim={1.2cm 2cm 3cm 1.5cm}, clip,width=0.9\linewidth]{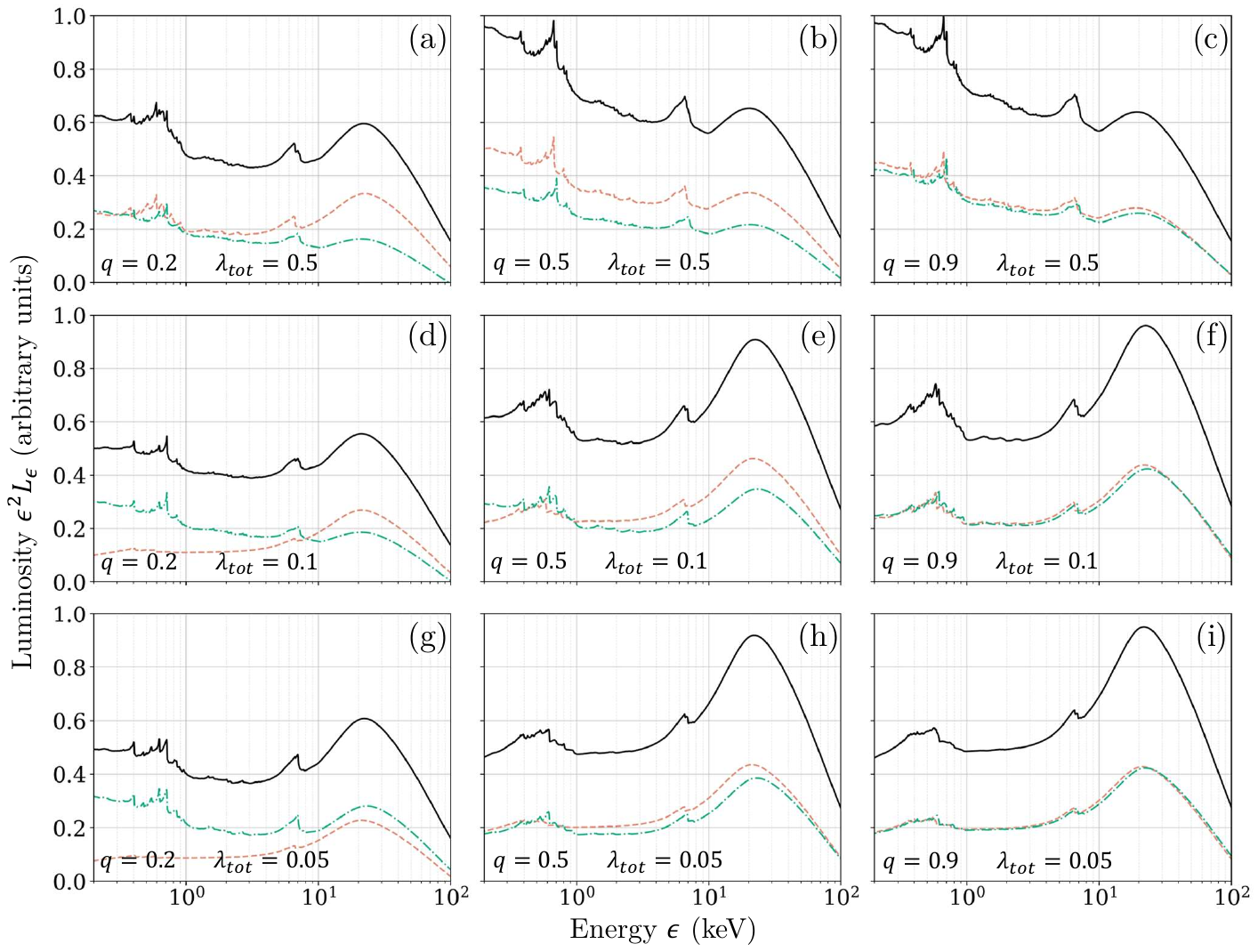}
    \caption{Total relativistic ionized reflection spectra from the 9 SMBH binary configurations shown in
      Table~\ref{tab:param} (solid lines). All are calculated for binaries in opposition ($f=90^{\circ}$) and the inclination of the observer $i=30^\circ$. The reflection spectra from the
      mini-disks around the primary and secondary SMBHs are shown as
      the dashed and dot-dashed lines, respectively. The y-axis has
      been scaled for each \ltot\ to show the spectra over the same
      relative range. When $q=0.9$, the two mini-disks produce very
      similar spectra which combine to accentuate the reflection
      features in the total spectrum. Unequal mass
      binaries are however impacted by the accretion inversion effect which
      causes the two mini-disks to have very different reflection
     properties. 
    \label{fig:grid}}
\end{figure*}

The final step before computing the total spectrum of a binary is to renormalize the individual spectra from each mini-disk to account for the relative sizes and accretion-powered luminosities of the disks. The \texttt{relxilllpCp} model produces spectra in units of flux defined per unit area of the emitting surface. Therefore, the relative contribution to the total luminosity of the secondary's mini-disk relative to that of the primary is defined as $\kappa\, (r_{\mathrm{out,2}}^2 -
    r_{\mathrm{in,2}}^2) / (r_{\mathrm{out,1}}^2 -
    r_{\mathrm{in,1}}^2)$
%
%
With the relative normalizations and velocity shifts now determined, the
composite X-ray spectra from each of the 9 binary configurations in
Table~\ref{tab:param} can be calculated at any orbital phase. 

%
%

\section{Composite X-ray Reflection Spectra of SMBH Binaries}
\label{sec:properties}

Figure~\ref{fig:grid} shows the relativistic ionized reflection
spectra from the mini-disks around the primary (dashed lines) and secondary (dot-dashed lines) SMBHs, \add{both modeled using \texttt{cflux(zashift*relxilllpCp)}}, 
along with the summed composite spectra (solid lines) from the 9
binaries listed in Table~\ref{tab:param}. The plotted spectra are calculated when each binary is at opposition
(i.e., $f=90^{\circ}$; Fig.~\ref{fig:geometry}), which maximizes the
line-of-sight velocity difference between the two SMBHs. As our focus
is to study the reflection properties of SMBH binaries, the effects
of ionized and neutral absorption, which are commonly observed in AGNs
\citep[e.g.,][]{ramos2017}, as well as any soft X-ray emission from a
warm corona \citep[e.g.,][]{ballantyne2020}, are omitted. 

Each row of Fig.~\ref{fig:grid} is scaled so that the effects of
increasing $q$ are easily seen for each \ltot. When $q=0.9$, the two
SMBHs are of near equal mass and are accreting at approximately the
same rates. As a result, both mini-disks are about the same size and at
very similar ionization states (Table~\ref{tab:param}). The resulting
reflection spectra from the two mini-disks are therefore nearly
identical (panels (c), (f) and (i) in Fig.~\ref{fig:grid}). This leads
to a total binary spectrum with more prominent reflection features,
with stronger emission lines and Compton hump (at $\approx 20$~keV)
than the spectra from either of the mini-disks. However, the composite
spectra are still dominated by features from a small range of
ionization parameters.

The effects of the accretion inversion phenomenon become stronger as
$q$ decreases. As $m_2$ and $m_1$ become more unequal, so do the
individual accretion rates, with the secondary accreting more rapidly
than the primary, leading to a more highly ionized disk (see
Table~\ref{tab:param}). When $q=0.2$ the peak ionization parameter of
the secondary can be well over an order of magnitude larger than the
primary (Fig.~\ref{fig:ionization_contour}). As a result, the
composite spectra of these binaries show features from a mix of
ionization states. Unlike the high $q$ case, the impact on the
composite can be limited to only a portion of the spectrum, depending
on the value of $q$ and \ltot. For example, in panels (a) and (e), the
soft X-ray emission lines below $\la 1$~keV are strongly affected
by the sum of different reflection features, while in panels (d) and (g), the impact is strongest for the Compton hump. These results show that
evidence for a SMBH binary may sometimes be found in only a part of an
observed X-ray spectrum, and not exclusively in the \fe\ line.

In general, a larger \ltot\ leads to a higher accretion rate and
therefore a more highly ionized mini-disk around both SMBHs (Fig.~\ref{fig:ionization_contour}). Because
of the accretion inversion effect, the mini-disk around the primary
is very weakly ionized when $\lambda_{\mathrm{tot}}=0.05$. At these
low ionization levels, the reflection spectrum is impacted by
absorption \citep{garcia2010}, and the predicted spectra from the
mini-disk are dominated by the contribution from the $\Gamma=2$
continuum, along with a relatively weak relativistically broadened
\fe\ line (see, e.g., panel (g) in Fig.~\ref{fig:grid}). At larger values of $\xi_1$ and
$\xi_2$, the reflection spectra from the mini-disks contain both a
relativistic \fe\ line from the inner disk and a number of soft X-ray
lines from larger radii due to the ionization gradient across the
disk. Crucially, even scenarios when the inner disk is nearly
completely ionized (e.g., the secondary mini-disk in panels (a)-(c) in
Fig.~\ref{fig:grid}),
the ionization gradient still produces an \fe\ line and soft X-ray
features from reflection at larger radii along the disk. Moreover, in most configurations shown in Fig.~\ref{fig:grid} the composite spectra exhibit visibly double-peaked \fe\ line profiles, which are of interest as a potentially unique signature of SMBH binaries.

\begin{figure*}[t!]
    \centering
    \includegraphics[width=0.95\linewidth]{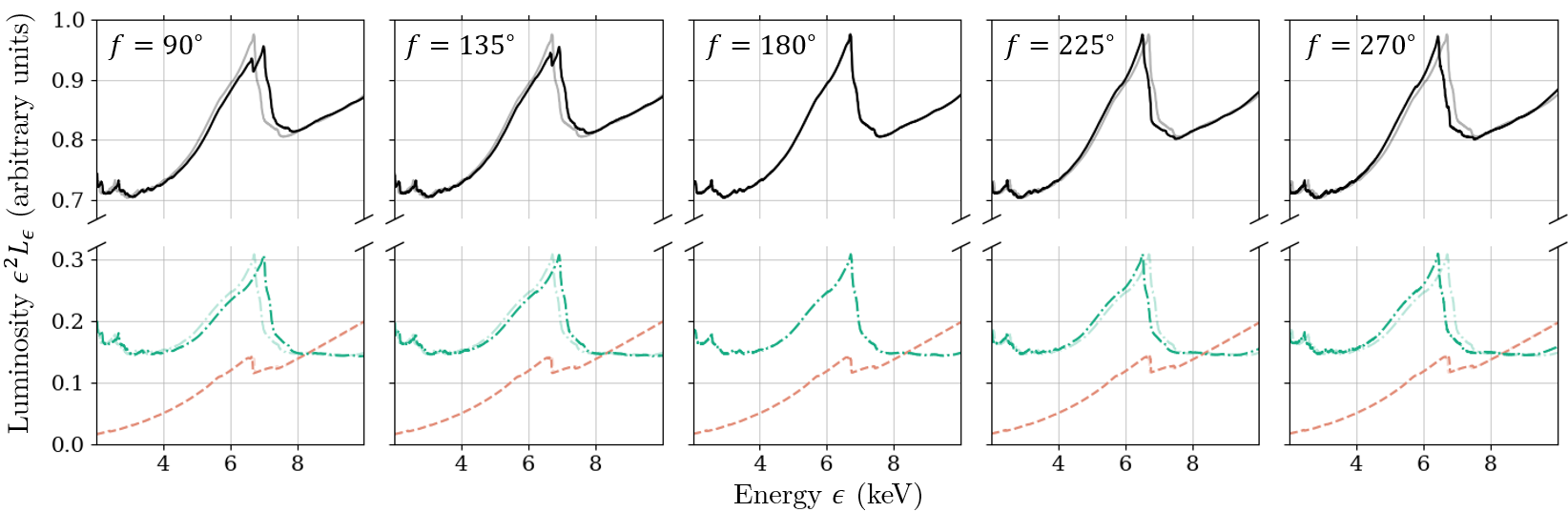}
    \caption{The effect of binary motion on the total \fe\ line profile from a SMBH binary at inclination $i=30\degree$ with $q=0.2$ and $\lambda_{\mathrm{tot}}=0.1$ (i.e., binary (d); Table~\ref{tab:param}). The line styles are the same as in Fig.~\ref{fig:grid}. The panels show the \fe\ line profile at $f=90^\circ$ (opposition; brighter secondary moving towards the observer), $f=135^\circ$, $f=180^\circ$ (conjunction),$f=225^\circ$  and $f=270^\circ$ (opposition; brighter secondary moving away from the observer). Each panel includes the profiles of the $f=0^\circ=180^\circ$ binary (with no orbital offset) as a semi-transparent curve for visual comparison. The relative motion of the two SMBHs leads to noticeable changes in the shape and offset of the \fe\ line profile.
    \label{fig:phaseshift}}
\end{figure*}

As mentioned above, Fig.~\ref{fig:grid} shows the SMBH reflection spectra at opposition ($f=90^{\circ}$) where there is the largest relative velocity between the two SMBHs. At the assumed distance of $100\,r_g$, the shift generates changes to the total binary spectra, especially around the \fe\ line and other narrow features in the soft end. The magnitude of the special relativistic Doppler shift for both SMBHs add up to $|z|=0.082$ for the assumed orbital inclination of $i=30^\circ$ (and is at most $|z|=0.095$ for edge-on orientation). Figure~\ref{fig:phaseshift} illustrates how the composite \fe\ line profile from binary (d) changes as a function of the orbital phase at $f=90^{\circ},~135^{\circ},~180^{\circ},~225^\circ$ and $270^{\circ}$. The modeled emission-line profile exhibits a visible change in shape, by evolving from double-peaked (for $f=90^\circ$) to a single-peaked profile. It also shows a shift of the blue edge of the emission line profile by about 0.5\,keV, consistent with the expected orbital Doppler shift due to the orbital motion of the brighter secondary.

%


In this example, the two mini-disks of the binary have very different ionization patterns. Due to the accretion inversion effect, the secondary SMBH is accreting rapidly and has a mini-disk that is very highly ionized at its inner edge. As a result, its \fe\ line comes from ionized iron at larger radii and the line profile is moderately broadened. In contrast, the primary SMBH is accreting weakly and the mini-disk is weakly ionized at the inner edge. Its reflection spectrum, therefore, exhibits a very relativistically broadened neutral (or weakly ionized) \fe\ line. The addition of this broad, low-contrast feature to the narrower ionized one from the mini-disk of the secondary produces modest changes to the composite line of the binary. However, the low ionization state of the primary's mini-disk gives its reflection spectrum a harder slope than the one from the secondary's mini-disk. Indeed, the spectrum from the primary SMBH dominates the composite at energies $>8$~keV. 

\section{Observational prospects}
\label{sec:prospects}
The previous section shows that the predicted reflection spectra of SMBH
binaries could be strongly impacted by the mass ratio $q$ and total
Eddington ratio \ltot\ of the binary. At large $q$, the total binary
spectrum will have strong reflection features due to the addition of
two very similar spectra from the mini-disks. However, when $q \la
0.5$, the individual reflection spectra may be very different due to
the accretion inversion effect, and the evidence of a binary may be
isolated to only certain regions of the total spectrum. In this Section, we
perform a limited investigation of the observational prospects for
detecting evidence of a SMBH binary from a typical long X-ray
observation. A more thorough study that considers a wider range of
binaries and observational strategies is deferred to a future paper.

\subsection{Setup}
\label{sub:setup}
\begin{figure}[t!]
    \centering
    \includegraphics[width=0.9\linewidth]{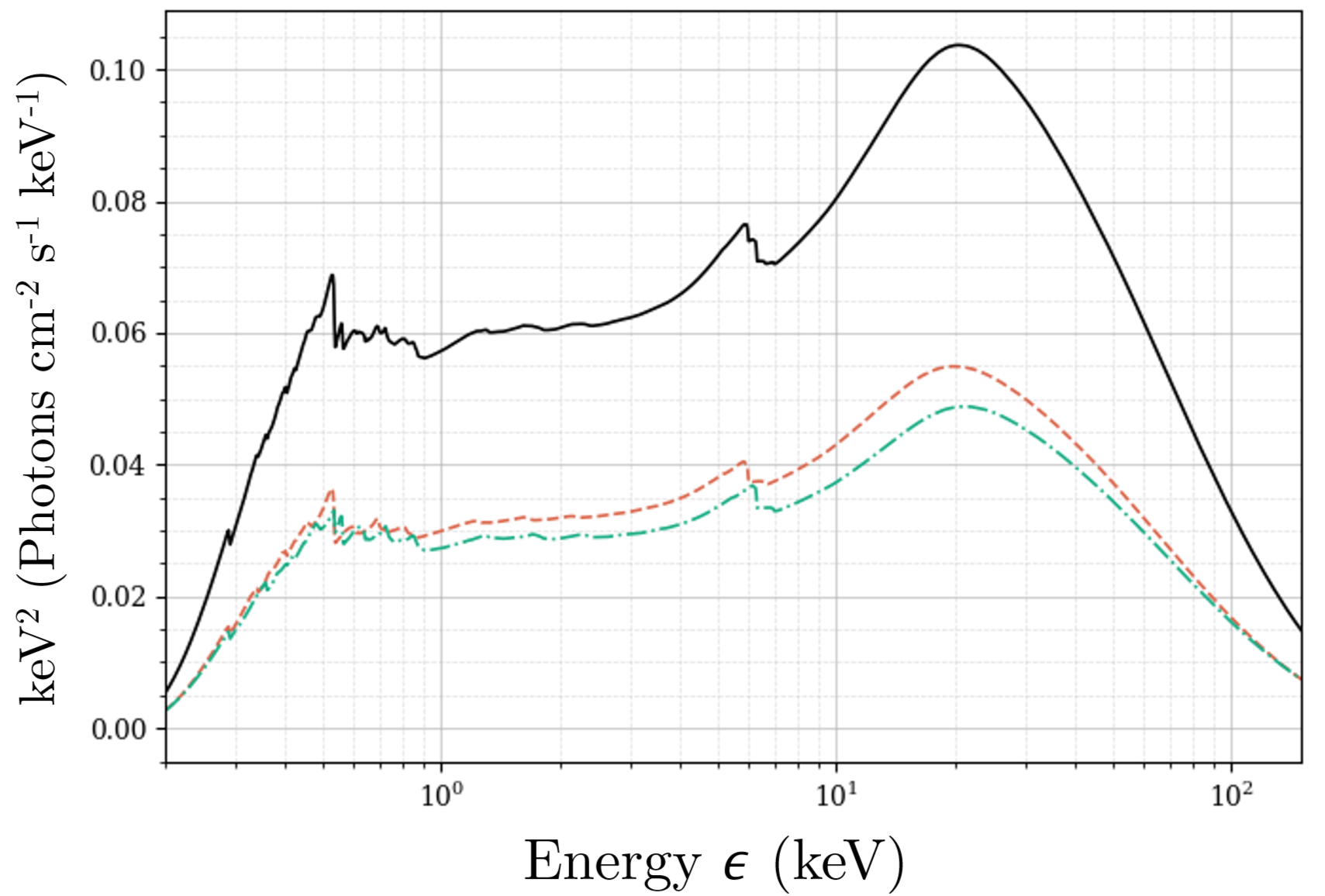}
    \caption{The X-ray spectral model used to simulate a 100~ks observation of a SMBH binary candidate. The line styles are the same as Fig.~\ref{fig:grid}. The binary considered has $q=0.9$ and $\lambda_{\mathrm{tot}}=0.1$ (i.e., binary (f); Table~\ref{tab:param}) and is shown at $z=0.1$ and at a phase angle of $90^{\circ}$. Galactic absorption with column density of $3\times 10^{20}$~cm$^{-2}$ is applied to the model, which was not the case for models from Sect.~\ref{sec:properties}. 
    \label{fig:spectrum}}
\end{figure}

We simulate 100~ks observations of the composite X-ray spectrum
from a SMBH binary with $q=0.9$ and
$\lambda_{\mathrm{tot}}=0.1$ (i.e., binary (f);
Table~\ref{tab:param}) at $z=0.1$ and at $z=1.0$. This configuration
is chosen because both mini-disks produce clear reflection signatures
across a wide range in energy. The total mass of
the binary is either $10^9$~M$_{\odot}$ or $10^6$~M$_{\odot}$,
referred to as a `PTA' or `\lisa' binary, respectively. The rest-frame
$2$--$10$~keV X-ray flux of each mini-disk is 
\begin{equation}
  \label{eq:fluxes}
  F_{\rm 2-10 keV} = f_X \frac{\lambda L_{\rm Edd}}{4 \pi d_L^2(z)},
\end{equation}
where $f_X$ is the \citet{duras2020} bolometric correction evaluated
at the appropriate Eddington ratio $\lambda$ for the mini-disk, and 
$L_{\mathrm{Edd}}$ is the Eddington luminosity for the primary or
secondary SMBH.  A $\Lambda$CDM cosmology with
$H_0=70$~km~s$^{-1}$~Mpc$^{-1}$, and $\Omega_m = 0.3$ is used to compute the
luminosity distance $d_{L}$ to redshift $z$. The resulting fluxes of
the primary and secondary mini-disks are then fixed using the
\texttt{cflux} command in \xspec.
Absorption from a Galactic column density of $N_{\mathrm{H}}=3\times
10^{20}$~cm$^{-2}$ is applied to the spectra of both mini-disks using
the \texttt{phabs} model. Figure~\ref{fig:spectrum} shows the model
spectrum for a $z=0.1$ PTA binary with $f=90^{\circ}$.
%
%


It is interesting to note that the exposure time (here adopted to be $100\,{\rm ks}\approx 1.2$\,days) is comparable to or larger than the orbital time of a \lisa\ binary with a separation close to $100$~$r_g$ (see equation~\ref{eq:torbital}). The resulting X-ray spectrum detected from such binaries would then effectively be orbit-averaged. To account for this effect, we average the model spectrum of
the \lisa\ binary over the full orbit to capture the smearing due to
the time-dependent orbital Doppler shifts. That is, we average over
the binary orbit's 4 extremal points: conjunction at $f=0\degree$ and
$180\degree$, and opposition at $f=90\degree$ and $270\degree$.
Conversely, the $100$~ks exposure time is much shorter than the
months-long orbital period of a PTA binary which results in a snapshot
of the spectrum at a fixed phase. The model for PTA binaries is set at
opposition ($f=90^\circ$) when the orbital shifts are most prominent
for both SMBHs. 

We simulate on-axis observations of the \lisa\ and PTA binaries with four
X-ray mission concepts that all had the potential to be operating at the same
time as \lisa:
\athena\ \citep{cruise2025},
\axis\ \citep{foord2024}, \hexp\ \citep{madsen2024},
and \strobex\ \citep{ray2024}. The latter two concepts both contain two instruments that observe
simultaneously and yield broadband spectra extending from below $0.5$~keV to past $30$~keV. While neither concept is currently being developed, it remains valuable to evaluate the potential usefulness of hard X-ray instruments in identifying SMBH binaries. \athena\ also has two instruments, but
observations are only made with one at a time, so we simulate separate
observations with either the Wide Field Imager (WFI) or the X-ray
Integral Field Unit (X-IFU). \axis\ and \athena\ are both sensitive
from $\approx 0.2$ to $\approx 12$~keV, and \strobex\ has the largest
collecting area of the four concepts, followed by \athena. Table~\ref{tab:instrument_files}
lists the specific response, ancillary and background files used to
produce our simulated observations from the four telescopes. The energy resolution of the simulated spectra are determined by the response files provided by each observatory. For reference, we also include Table~\ref{tab:instrument_capabilities}, where we indicate the nominal bandpass and collecting area at 6~keV of each instrument.  All
simulations are performed with \texttt{fakeit} in \xspec\ and the
resulting spectra are grouped to have a minimum of 20 counts per bin.

%
\begin{figure*}[t!]
    \centering
    \includegraphics[width=0.9\linewidth]{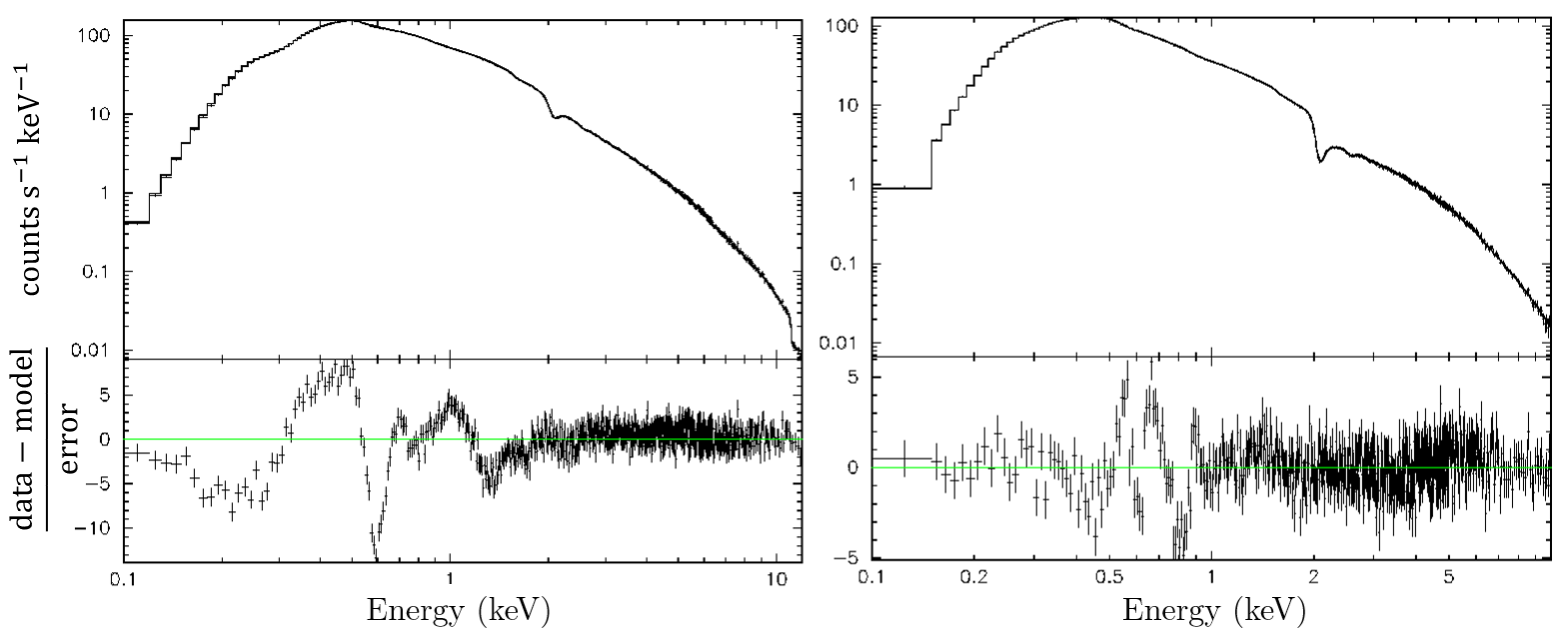}
    \caption{Left: The upper-panel shows the predicted count rate
      spectrum of a $z=0.1$ PTA SMBH binary (with total mass
      $M=10^9$~M$_{\odot}$, $q=0.9$ and $\lambda_{\mathrm{tot}}=0.1$)
      observed by \athena\ (WFI) for $100$~ks. The lower-panel shows
      the residuals when this spectrum is fit by a spectral model for a
      single AGN (\texttt{phabs*relxill}). Strong soft X-ray residuals
      are present due to the combination of emission features from the two
      mini-disks. Right: As in the left-hand panel but now showing
      the result from the simulated \axis\ observation.
    }
    \label{fig:fits}
\end{figure*}

Our goal is to emulate the process of fitting a spectrum from an
unknown AGN under the hypothesis that it is emission from a single
SMBH, where the observer knows only the Galactic column density in the
direction of the source and the redshift of the object. Therefore, the simulated spectra are fit in \xspec\ using a \texttt{phabs*relxill}
model with the column density fixed at $3\times 10^{20}$~cm$^{-2}$ and
the redshift is set to either $0.1$ or $1$. Within
\texttt{relxill}, the disk emissivity\footnote{The standard model for \texttt{relxill} does not make any pronouncements as to the nature or geometry of the corona, and uses a simple empirical power-law to model disk emissivity.} is set to $-3$, the inner disk radius is fixed at the ISCO, the
outer radius is frozen at $400$~$r_g$  and the inclination
angle is set to $30\degree$. The remaining
\texttt{relxill} parameters (SMBH spin $a$, photon index $\Gamma$,
ionization parameter $\log \xi$, iron abundance $A_{\mathrm{Fe}}$ and
reflection fraction $f_{\mathrm{refl}}$) are all left free to vary
when fitting the simulated spectra. Goodness of fit is determined
using $\chi^2$ statistics, and uncertainties on the fit parameters are
reported using a $\Delta \chi^2=2.71$ criterion, or a 90\% confidence interval for one parameter of interest \add{using the \texttt{error} command in \xspec}. Prior to fitting, we
ignore any channels that are background dominated. Therefore, some
fits, especially those for models $z=1$, are performed over a narrower
energy range than the native energy sensitivity of the instrument.

\subsection{Fit Results}
\label{sub:fits}

We examine the fit results to search for evidence of the SMBH binary
in the simulated spectrum. This evidence could manifest as a poor fit (i.e., a
$\chi^2/$d.o.f. $> 2$, where d.o.f. = degrees of freedom), or the fit could be acceptable, but one or more
parameters may yield an unusual or unexpected result. A third
possibility is that the fit and parameters may be reasonable, but
residuals may remain that could be difficult to explain using
conventional AGN phenomenology. We are also interested in determining if
one of the 4 X-ray observatories is best suited to finding evidence
for SMBH binaries.

%
\begin{deluxetable*}{ccccccc}
\tabletypesize{}
\tablewidth{0pt} 
\tablecaption{Best fit parameters when a \texttt{phabs*relxill} model
  is applied to a simulated PTA binary spectrum at opposition ($M=10^9$~M$_{\odot}$, $q=0.9$, $\ltot = 0.1$, and $f=90^{\circ}$). The
  composite binary spectrum is subject to Galactic absorption
  ($N_{\mathrm{H}}=3\times 10^{20}$~cm$^{-2}$) and placed at either
$z=0.1$ or $1$ (Fig.~\ref{fig:spectrum}). Individual faked spectra are
generated for 100~ks exposures by \athena\ (WFI), \athena\ (X-IFU),
\axis, \hexp, and \strobex. The free parameters when fitting the
\texttt{relxill} model are the spin parameter $a$, photon index
$\Gamma$, ionization parameter $\text{log}~\xi$, iron abundance (in
Solar units) and the reflection fraction
  $f_{\text{refl}}$. The redshift and absorbing column density are
fixed at the known values. Uncertainties
  correspond to a 90\% confidence range on the fit parameter  \add{using the \texttt{error} command in \xspec}. A `p'
  in an error bar indicates that the error range pegged at the upper
  or lower limit of the parameter.\label{tab:pta}} 
\tablehead{
\colhead{$z$} & \colhead{$\chi^2 \, /$ d.o.f.} & \colhead{$a$}& \colhead{$\Gamma$} & \colhead{$\text{log}~\xi$} & \colhead{A$_{\rm Fe}$} & \colhead{$f_{\text{refl}}$}
} 
\startdata
\multicolumn{7}{c}{\athena\ (X-IFU)} \\
\cline{1-7} 
\\[-5pt]
{$0.1$} & {15,816 / 12,164} &{$0.998\substack{+0.000p\\-0.001}$} & {$1.95 \pm 0.00$} &{$2.46 \pm 0.01$}&{$0.88 \pm 0.02$}&{$0.86 \pm 0.01$} \\
{$1.0$} & {1,183 / 1,235} & {$0.997\substack{+0.001p \\ -0.633}$} & {$1.90 \substack{+0.03 \\ -0.04}$} & {$2.90\substack{+0.26 \\ -0.22}$} & {$0.66\substack{+0.24 \\ -0.16p}$} & {$1.08\substack{+1.32 \\ -0.39}$} \\ \\[-5pt]
\cline{1-7}
\multicolumn{7}{c}{\athena\ (WFI)} \\
\cline{1-7} 
\\[-5pt]
{$0.1$} & {3,958 / 1,137} & {$0.998$} & {$1.94$} &{$2.44$}&{$0.85$}&{$0.90$} \\
{$1.0$} & {375 / 423}     & {$0.323\substack{+0.675p \\ -0.567}$} & {$2.00\substack{+0.03 \\ -0.01}$} & {$1.77\substack{+0.18 \\ -0.77}$} & {$1.92\substack{+1.22 \\ -1.12}$} & {$1.17\substack{+0.40 \\ -0.26}$} \\ \\[-5pt]
\cline{1-7}
\multicolumn{7}{c}{\axis} \\
\cline{1-7}
\\[-5pt]
{$0.1$} & {1,304 / 893} &{$0.972\substack{+0.016\\-0.018}$} & {$2.01 \pm 0.00$} &{$1.63 \pm 0.02$}&{$0.76\substack{+0.009\\-0.011}$}&{$1.22 \pm 0.04$} \\
{$1.0$} & {332 / 360} & {$0.524\substack{+0.474p \\ -1.522p}$} & {$2.00 \substack{+0.03\\-0.04}$} & {$1.73\substack{+0.66 \\ -0.92}$} & {$0.50\substack{+3.41 \\ -0p}$} & {$0.92\substack{+1.19 \\ -0.36}$} \\ \\[-5pt]
\cline{1-7}
\multicolumn{7}{c}{\hexp} \\
\cline{1-7}
\\[-5pt]
{$0.1$} & {2,586 / 2,381} &{$0.998\substack{+0.000p\\-0.010}$} & {$1.94\pm 0.00$} &{$2.51 \pm 0.04$}&{$0.84 \pm 0.05$}&{$0.96 \pm 0.02$} \\
{$1.0$} & {160 / 201} & {$0.667\substack{+0.331p \\ -1.665p}$} & {$1.85 \pm 0.07$} & {$3.10\substack{+0.21 \\ -0.36}$} & {$0.52\substack{+0.38 \\ -0.02p}$} & {$2.65\substack{+7.35p \\ -1.71}$} \\ \\[-5pt]
\cline{1-7}
\multicolumn{7}{c}{\strobex} \\
\cline{1-7} 
\\[-5pt]
{$0.1$} & {2,953 / 1,621} & {$0.987\substack{+0.011p \\ -0.010}$} & {$2.01 \pm 0.00$} & {$1.64\substack{+0.01 \\ -0.02}$} & {$0.91 \pm 0.03$} & {$1.35 \substack{+0.02 \\ -0.03}$} \\
{$1.0$} & {535 / 536} & {$0.599\substack{+0.399p \\ -1.597p}$} & {$1.91 \substack{+0.03 \\ -0.04}$} & {$2.92\substack{+0.30 \\ -0.34}$} & {$0.50 \substack{+0.19 \\ -0p}$} & {$0.93 \substack{+9.07p \\ -0.25}$} \\ \\[-5pt]
\enddata
\end{deluxetable*}

The fit results for our simulated PTA ($10^9$~M$_{\odot}$) binaries
are shown in Table~\ref{tab:pta}. As a reminder, the models used in
the simulations all have $\Gamma=2$, $A_{\mathrm{Fe}}=1$ in solar units and $a=0.99$. 
There is a significant distinction in the results between a source at
$z=0.1$ (with a $2$--$10$~keV flux of $F=2.61 \times 10^{-11}$~erg~cm$^{-2}$~s$^{-1}$) and one at
$z=1$ ($F=1.28 \times 10^{-13}$~erg~cm$^{-2}$~s$^{-1}$). All 5 simulated observations of the
$z=1$ binary are well fit with the simple single AGN model
indicating that the effects of the two reflection signals are not
leading to any significant residuals. However, there are hints in the
values of the fit parameters that there are aspects of the simulated
spectra that are not well described with one \texttt{relxill}
model. In particular, the fitted SMBH spin $a$ is almost entirely
unconstrained in all 5 observations. Only the X-IFU observation gets
close to the `true' value of $a=0.99$, but with a large negative
error-bar. This result indicates that shape of the \fe\ line in the
model spectrum does not match that expected from a single AGN and the
fit becomes insensitive to its spin value.

The high energy sensitivity of \hexp\ and \strobex\ also provide some
clues for the presence of a binary at $z=1$. In both cases, the fits
return a harder $\Gamma$ than the input value, along with a low
iron abundance and a large and poorly constrained reflection fraction
($f_{\mathrm{refl}}$). \add{This seems to suggest that, with their energy coverage beyond $\sim 12$~keV, these observatories would be better positioned to capture the enhanced Compton reflection hump due to}
 the combination of two reflection signals (Fig.~\ref{fig:spectrum}). \add{This may help in resolving some degeneracies between the parameters responsible for determining the relative strength of reflection features.}
Of the other observatories, only the X-IFU
observation gives a hint of a harder spectrum, although both
$A_{\mathrm{Fe}}$ and $f_{\mathrm{refl}}$ return values typical of a
single AGN. The large soft X-ray sensitivity of \axis\ allows it to
find the correct $\Gamma$, but the fit value for iron abundance remains largely
unconstrained\add{, as with the WFI}. 

The results for a PTA binary at $z=0.1$
are dramatically different than those at $z=1$ as all the simulated observations (except for
\hexp) cannot be adequately fit by a single AGN
model\footnote{We do note that the \hexp\ fit still shows clear residuals mostly clustered in the soft end similarly to the other instrument fits, despite a $\chi^2$ value of $\sim 1.1$.}. In fact, the \athena\ (WFI) fit is so poor (with
$\chi^2/$d.o.f.$> 2$) that error-bars cannot be calculated. The reason for the poor fits is illustrated in
Figure~\ref{fig:fits} which shows the results from the mock
\athena\ (WFI) and \axis\ observations. In both cases, strong residuals are seen between $\approx 0.5$ and
$0.8$~keV, indicating that the single AGN model cannot account
for the combination of spectral features predicted in the
composite binary spectrum. The complexity of
the soft X-ray line emission predicted by the model arises from ionization
gradients on both mini-disks as well as the small velocity shifts
between the two SMBHs. As a result, a model that predicts the
reflection spectrum from one accretion disk with a single ionization
parameter is unable to reproduce the spectrum that our model predicts
from the binary. Although the fits of the $z=0.1$ PTA binary are almost all
statistically poor, the resulting fit parameters are all well
constrained and are close to the `true' values (e.g., the SMBH
spin). Interestingly, even with these high signal-to-noise spectra,
the composite \fe\ line predicted from the binary does not produce
significant residuals when fit with a single AGN model. Therefore, if
they can be convincingly separated from any warm absorption or coronal emission,
complex soft X-ray emission lines may be the most direct evidence for
a SMBH binary in a single X-ray spectrum. 

%
%
\begin{deluxetable*}{ccccccc}
\tabletypesize{}
\tablewidth{0pt} 
\tablecaption{As in Table~\ref{tab:pta}, but now assuming a
  \lisa\ binary ($M=10^6$~M$_{\odot}$). In this case, the
  composite SMBH binary spectrum is averaged over four phases of
  the orbit ($f=0^{\circ}$, $90^{\circ}$, $180^{\circ}$, and
  $270^{\circ}$) since the 100~ks exposure time is comparable to the
  orbital period of the binary. \label{tab:lisa}}
\tablehead{
\colhead{$z$} & \colhead{$\chi^2 \, /$ d.o.f.}  & \colhead{$a$}& \colhead{$\Gamma$} & \colhead{$\text{log}~\xi$} & \colhead{A$_{\mathrm{Fe}}$} & \colhead{$f_{\text{refl}}$}
} 
\startdata
\multicolumn{7}{c}{\athena\ (X-IFU)} \\
\cline{1-7} \\[-8pt]
{$0.1$} & {76 / 89} & {$0.926\substack{+0.072p \\ -1.703}$} &
{$2.09\substack{+0.08 \\ -0.14}$} & {$1.31\substack{+0.65 \\ -0.84}$}
& {$8.89\substack{+1.11p \\ -8.39p}$} & {$4.11\substack{+3.17
    \\ -2.91}$} \\
[+5pt]
\cline{1-7}
\multicolumn{7}{c}{\athena\ (WFI)} \\
\cline{1-7} \\[-8pt]
{$0.1$} & {166 / 150} & {$0.498\substack{+0.500p \\ -1.496p}$} &
{$2.00\substack{+0.10 \\ -0.17}$} & {$2.30\substack{+1.70 \\ -2.19}$}
& {$0.50\substack{+9.50p \\ -0p}$} & {$0.46\substack{+1.79 \\ -0.38}$}
\\
[+5pt]
\cline{1-7}
\multicolumn{7}{c}{\axis} \\
\cline{1-7} \\[-8pt]
{$0.1$} & {109 / 98} & {$-0.180\substack{+1.178p \\ -0.818p}$} &
{$2.03\substack{+0.55 \\ -0.94}$} & {$1.92\substack{+2.18 \\ -1.92p}$}
& {$0.50\substack{+9.50p \\ -0p}$}  & {$2.54\substack{+7.46p
    \\ -2.43}$}\\
[+5pt]
\cline{1-7}
\multicolumn{7}{c}{\hexp} \\
\cline{1-7} \\[-8pt]
{$0.1$} & {29 / 28} & {$-0.139\substack{+1.137p \\ -0.859p}$} &
{$1.58\substack{+0.29 \\ -0.24}$} & {$3.22\substack{+0.85 \\ -1.34}$}
& {$6.97\substack{+4.92 \\ -3.03p}$} & {$9.70\substack{+0.30p \\ -8.86}$} \\
[+5pt]
\cline{1-7}
\multicolumn{7}{c}{\strobex} \\
\cline{1-7} \\[-8pt]
{$0.1$} & {150 / 154} & {$0.989\substack{+0.009p \\ -1.987p}$} & {$1.10\substack{+0.48 \\ -0.10p}$} & {$3.54\substack{+1.16p \\ -0.47}$} & {$5.11\substack{+4.89p \\ -3.85}$} & {$10.00\substack{+0.00p \\ -10.00p}$} \\
[+5pt]
\enddata
\tablecomments{At redshifts $z=1.0$, although sources are detectable
  and their spectral shape potentially identifiable as
  characteristically AGN-like, the predicted count rates are
  insufficient for spectral analysis and $\chi^2$ statistics with a
  $100$~ks exposure.}
\end{deluxetable*}

Fit results for simulated observations of a \lisa\ binary
($M=10^6$~M$_{\odot}$) are shown in Table~\ref{tab:lisa}. This
source is $10^3$ times fainter than the PTA binary ($2$--$10$~keV flux of $F=2.61 \times 10^{-14}$~erg~cm$^{-2}$~s$^{-1}$ at $z=0.1$ and $F=1.28 \times 10^{-17}$~erg~cm$^{-2}$~s$^{-1}$ at $z=1$) and we find that
spectral analysis is impossible for such a binary at $z=1$ with a
$100$~ks exposure from any of the 4 observatories. The predicted
number of counts does not give enough channels above background to
perform more than a rudimentary fit. While such a source is likely
detectable by all the observatories and potentially identified as an
AGN, detailed spectral analysis can only be done through significantly
longer exposures. If a \lisa\ binary is placed at $z=0.1$, then
$100$~ks exposures are sufficient to perform spectral fitting for all
instruments. The results are broadly consistent with those found from
the $z=1$ PTA binary discussed above: fits using a single AGN spectral
model are all successful, the SMBH spin parameter is unconstrained,
and \hexp\ and \strobex\ show evidence of a very hard spectrum with a
large iron abundance. It is notable that even with the large
collecting area of \athena, exposure times $\gg 100$~ks may be needed
to convincingly uncover evidence of SMBH binaries destined to be
\lisa\ sources. 

\subsection{Detecting Orbital Variability in \fe\ Line Profiles}
\label{sub:variability}
The above discussion showed that a single X-ray observation of a
source may not be sufficient to find clear evidence of a SMBH
binary. However, for high mass PTA binaries, multiple X-ray
observations of the same object will probe different phases of the binary orbit, and the
resulting changes in the composite spectrum could be identified
through spectral fitting. As shown in Fig.~\ref{fig:phaseshift},
changes in the \fe\ line profile with orbital phase could be an
important indicator of the presence of a SMBH binary.

We test this possibility by simulating 100~ks
\athena\ observations of the PTA binary shown in Fig.~\ref{fig:phaseshift}
(i.e., $M=10^9$~M$_{\odot}$, $q=0.2$ and $\lambda_{\mathrm{tot}}=0.1$)
at phases of $f=90^{\circ}$, $180^{\circ}$ and $270^{\circ}$. The
three model binary spectra are computed as in Sect.~\ref{sub:fits} and
the source is assumed to be at $z=0.1$. Given the high
count-rate predicted for the \athena\ observations, we group the mock
observations to have a minimum of 50 counts per bin. As our goal is to
determine if the different orbital phases can be detected in the
composite \fe\ profile, we fit the mock spectra between $2$ and
$12$~keV with a \texttt{phabs*relxill} spectral model. This energy
range eliminates the possibility of soft X-ray features dominating the fitting procedure.

%
%
\begin{deluxetable*}{ccccccc}
\tabletypesize{}
\tablewidth{0pt} 
\tablecaption{As in Table~\ref{tab:pta}, but now fitting simulated
  \athena\ spectra of a $q=0.2$ PTA binary at $z=0.1$ observed at
  three different phases (Fig.~\ref{fig:phaseshift}). The goal of these fits is to
  search for orbital effects around the \fe\ line so all fits are
  performed between $2$ and $12$~keV. Due to the higher predicted
  count rates, the simulated spectra are re-grouped so that each bin
  contains at least 50 counts. \label{tab:variability}}
\tablehead{
\colhead{$f$} & \colhead{$\chi^2 \, /$ d.o.f.} & \colhead{$a$}& \colhead{$\Gamma$} & \colhead{$\text{log}~\xi$} & \colhead{A$_{\rm Fe}$} & \colhead{$f_{\text{refl}}$}
} 
\startdata
\multicolumn{7}{c}{\athena\ (X-IFU)} \\
\cline{1-7} 
\\[-5pt]
{$90$}  & {5,828 / 5,920} & {$ 0.454\substack{+0.279 \\ -1.452p}$} & {$1.95 \substack{+0.01 \\ -0.03}$} & {$3.01 \substack{+0.28 \\ -0.13}$} & {$0.53\substack{+0.08 \\ -0.03p}$} & {$0.55 \substack{+0.14 \\ -0.10}$} \\ 
{$180$} & {5,886 / 5,933} & {$ 0.610\substack{+0.181  \\ -0.434}$} & {$1.96 \pm 0.02$}                  & {$3.00 \substack{+0.07 \\ -0.14}$} & {$0.57\substack{+0.15 \\ -0.07p}$} & {$0.72 \substack{+0.23 \\ -0.14}$} \\
{$270$} & {5,886 / 5,933} & {$ 0.605\substack{+0.157  \\ -0.225}$} & {$2.11 \pm 0.03 $}                 & {$1.38 \substack{+0.25 \\ -0.80}$} & {$0.87\substack{+0.11 \\ -0.10}$}  & {$1.49 \substack{+0.27 \\ -0.25}$} \\ \\[-5pt]
\cline{1-7} 
\multicolumn{7}{c}{\athena\ (WFI)} \\
\cline{1-7} 
\\[-5pt]
{$90$}  & {842 / 836} & {$-0.998\substack{+0.783 \\ -0.000p}$} & {$1.91 \substack{+0.01 \\ -0.02}$} & {$3.28 \substack{+0.07 \\ -0.09}$} & {$0.51\substack{+0.08 \\ -0.01p}$} & {$0.62 \substack{+0.70 \\ -0.13}$} \\ 
{$180$} & {842 / 834} & {$ 0.617\substack{+0.173  \\ -0.172}$} & {$1.94 \substack{+0.04 \\ -0.02}$} & {$3.06 \substack{+0.12 \\ -0.22}$} & {$0.71\substack{+0.25 \\ -0.13}$}  & {$0.71 \substack{+0.24 \\ -0.15}$} \\
{$270$} & {777 / 833} & {$ 0.992\substack{+0.006p \\ -0.493}$} & {$1.95 \substack{+0.01 \\ -0.04}$} & {$3.08 \substack{+0.14 \\ -0.22}$} & {$0.61\substack{+0.05 \\ -0.11}$}  & {$1.29 \substack{+0.18 \\ -0.56}$} \\ \\[-5pt]
\enddata
\end{deluxetable*}

Table~\ref{tab:variability} shows the results for both X-IFU and WFI
observations of the binary. For both instruments, the spectra from all
three orbital phases are well fit by the single AGN spectral
model. 
The largest spectrum-to-spectrum change is seen from the simulated X-IFU observation of the
$f=270^{\circ}$ model which yields a significantly different ionization
parameter and a higher reflection fraction than for the other orbital
phases. However, these types of variations in reflection parameters
are not uncommon when fitting multiple observations of a single AGN
\citep[e.g.,][]{keek16}, and can be interpreted as arising from the
complex dynamical environment of an accretion disk corona rather
than evidence for a binary. Similar to what was found with
our previous simulations, all of the fits struggled to constrain the
SMBH spin with values often changing drastically between orbital
phases. Notably, the WFI observation of the $f=90^{\circ}$ spectrum
returns a negative spin parameter, while the fits to the other two
phases are consistent with a typical prograde spin. Repeated X-ray observations of a source
that exhibits such variability in its measured SMBH spin could indicate the
presence of an unusual and changing \fe\ line profile consistent with the presence of a SMBH binary (e.g., Fig.~\ref{fig:phaseshift}).


\begin{figure*}[t!]
    \centering
    \includegraphics[width=0.49\linewidth]{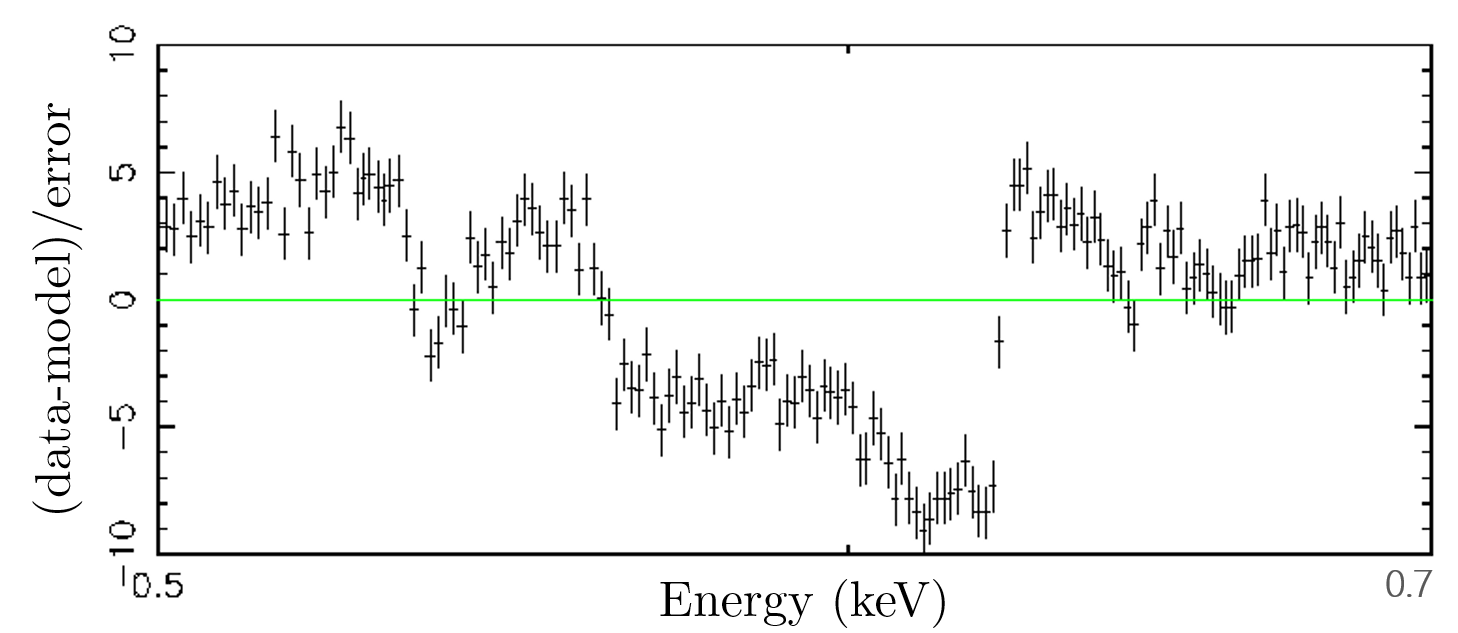}
        \includegraphics[width=0.49\linewidth]{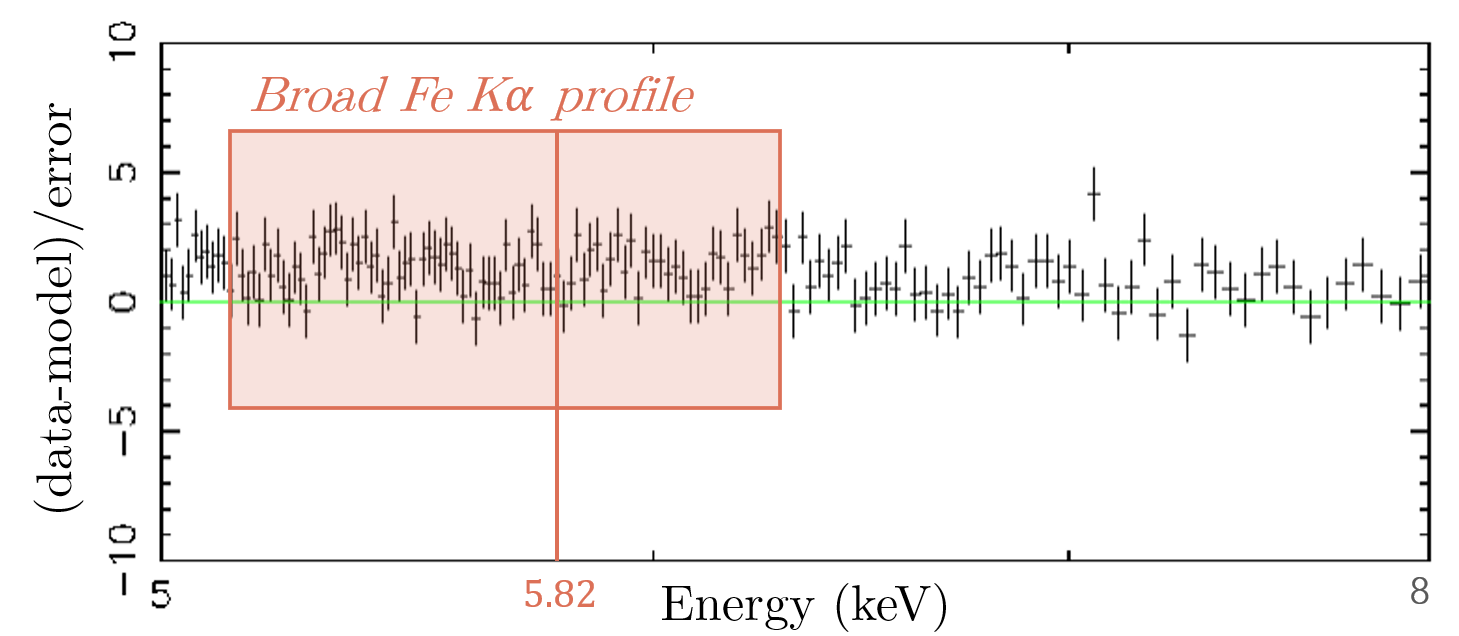}
    \caption{Left: The $0.5$ to $0.7$~keV residuals found after fitting the 100~ks mock X-IFU observation of a $z=0.1$ PTA SMBH binary with a single AGN model (see Table~\ref{tab:pta}). The high spectral resolution of the X-IFU reveals the complex combination of lines and emission features that are predicted in the composite SMBH binary model (Fig.~\ref{fig:spectrum}). Right: As in the left panel, but now focused on the energy band that encompasses the \fe\ line (around 5.8 to 6.0 keV at z=$0.1$). The orange box highlights the extent of the broad \fe\ line profile; a line marks its rest energy, $6.4/(1+z) =5.82$~keV at $z=0.1$. While the X-IFU cannot resolve the small double-peak predicted by the composite spectrum,  the clear excess flux between $5$ and $6$~keV, as evidenced by the consistently positive residuals, is consistent with the presence of a second \fe\ line in the spectrum.
\label{fig:xifu}}
\end{figure*}

\section{Discussion}
\label{sec:discussion}
\subsection{Feasibility of SMBH Binary Detection Through X-ray Reflection Signatures}
\label{sub:detection}

Our analysis points to several outcomes relevant to detection of SMBH binaries using X-ray reflection spectroscopy. One is that by themselves, the single-epoch X-ray spectra are not likely to be a smoking gun that can be used to decisively diagnose a binary. They can however provide supporting evidence for it by indicating that a single AGN model does not provide a satisfactory fit to the data. This case can be most convincingly made for the nearest and most luminous SMBH binaries, where the single AGN model fit fails in the most obvious way: by having a poor $\chi^2$ statistic\add{ despite the high signal-to-noise ratio. Although such discrepancies are not uncommon in AGN spectral fitting, they warrant closer scrutiny of the reasons behind the poor fits and may in turn provide supporting evidence for the binary hypothesis}. In the case of more distant ($z=1$) or less luminous SMBH binaries (\lisa\ sources) that are detected with a sufficient number of counts for a fit to be attempted, the case is more subtle. In such scenarios we find that the quality of the spectral fit may be adequate from a statistical ($\chi^2$) standpoint but the value of the SMBH spin tends to be unconstrained. In addition to the anomalous spin measurements, in such cases the fits of the spectra with sufficient high energy coverage also tend to indicate relatively hard X-ray spectra, and sometimes (but not consistently) a large or poorly constrained abundance and reflection fraction. 

In most cases included in our analysis, the \fe\ lines in one-epoch spectra are fit reasonably well with a single AGN model and do not show significant fit residuals. Even so, they seem to contain useful information about the spin of one or both SMBHs that can potentially be recovered from fit. Specifically, we find that in the case of the nearby and luminous SMBH binaries with $q\approx 1$, the single AGN model returns the spin value representative of the input spin we assigned to the two SMBHs in our model. While the universality of this outcome remains to be fully explored for a wide range of binary configurations (and in particular those with different values of the primary and secondary spins), it is intriguing as it may provide a way to place constraints on the spin of one or both SMBHs. This is of particular importance for the PTA-detected binaries, as GW parameter estimation in these systems is not expected to place constraints on the SMBH spins and the X-ray reflection spectroscopy may provide a unique opportunity to do so. 

The nearby luminous X-ray sources associated with PTA-like massive SMBH binaries have one more important advantage: \add{that} X-ray spectroscopy can be used to probe different phases of the binary orbit, providing the opportunity to investigate the time-variable binary signatures. In the example of a $q=0.2$ binary the time-variability effect is most pronounced in the spin measurement. Although the input spin of neither SMBH changes from one point on the orbit to another, the phase-dependent variation in the shape of the \fe\ emission-line profile causes the single AGN model to return widely varying best fit values. This evidence, reinforced by a substantial variation in the ionization parameter and the reflection fraction would point to unusual and possibly unique physical conditions in the nuclear accretion flow of the emitting source. Note that as a consequence of the widely varying inferred spin value in the case of the $q=0.2$ SMBH binary, the fit does not consistently recover the input spin we assigned to the SMBHs. This is in apparent contrast with the $q\approx 1$ case discussed in the previous paragraph and it motivates the development of a more rigorous parameter estimation method for SMBH binaries that is beyond the scope of this study. 

\subsection{Binary Identification Capabilities of Future X-ray Missions}

Our analysis indicates that all X-ray observatories examined here are to some degree sensitive to the fact that the composite SMBH binary X-ray spectrum cannot be optimally fit by a model of a single AGN. In the next few paragraphs we discuss the most effective observational strategies for the X-ray detection and follow-up of SMBH binaries that are GW sources.

The combination of large collecting area (0.60 m$^2$ at 1~keV) and high spectral resolution ($\leq 4$~eV at 7~keV) makes the X-IFU instrument on \athena\ a powerful tool for identifying SMBH binaries \citep{cruise2025}. This is demonstrated in Figure~\ref{fig:xifu} which shows the fit residuals from a mock X-IFU observation of the PTA binary at $z=0.1$ (see also Table~\ref{tab:pta}). The left panel shows the detailed residuals in the $0.5-0.7$\,keV band that result when fitting this spectrum with the single AGN model. The mix of soft X-ray recombination lines from two mini-disks with a velocity offset yields a rich series of residuals in the high-resolution X-IFU spectrum. The prominent soft X-ray residuals can also be identified in the simulated \axis\ observation in the right panel of Figure~\ref{fig:fits}. While any ionized absorption present along the line-of-sight (not accounted for in our model) will undoubtedly complicate the analysis, the features predicted to be present in the soft X-ray part of the \athena\ X-IFU and \axis\ spectra may be prominent enough to reveal a SMBH binary.

The right panel of Fig.~\ref{fig:xifu} shows the residuals around the \fe\ line when the X-IFU spectrum is fit with the single AGN model. The assumed 100~ks exposure time is not sufficient for the high spectral resolution of the X-IFU to detect the double-peaked composite \fe\ line profile due to the binary (Fig.~\ref{fig:spectrum}). However, the residuals clearly show a positive excess between $5$ and $6$~keV resulting from the sum of the two red-wings in the \fe\ lines from both mini-disks. This significant excess, in concert with the large soft X-ray residuals illustrated in the other panel, could be compelling evidence for the presence of a SMBH binary.

\add{Conversely, observatories with the most reliable high energy X-ray coverage (\hexp\ and \strobex) seem to more often produce spectra with anomalous best-fit values for the photon index, iron abundance and/or reflection fraction.}
\add{By capturing the true extent of the Compton hump, these instruments may be able to disentangle degeneracies between these parameters and recognize the inherent discrepancies of a composite spectrum.}

The poor $\chi^2$ statistic, unconstrained spin and unusual values of other parameters discussed above would be much more suggestive of the binary presence if a GW detection of an inspiraling SMBH binary was announced in the same part of the sky. In such cases, the X-ray spectroscopy could be carried out as a part of the EM follow-up of a GW source. The GW localization regions of PTA detected binaries are however expected to be relatively large \citep[corresponding to $\sim 10{\rm s}-100{\rm s}\,{\rm deg^2}$ on the sky;][]{petrov24}, compared to the fields of view of the X-ray instruments considered in this work (which tend to be comparable to or less than $0.4\,\rm deg^2$). It follows that the most efficient way to obtain the X-ray spectroscopy would be after a precise location of the EM counterpart was determined from the archival data or by another instrument like the Wide Field Monitor (WFM) on \strobex, for which the instantaneous field of view covers about 32\% of the sky.  As noted before, SMBH binaries with orbital separations of $100\,r_g$ constitute \lisa\ precursor sources and are only going to be detected by it later in the inspiral and at merger. The GW localization region of \lisa-detected binaries is expected to reach $\sim 0.1\,{\rm deg^2}$ at merger \citep{mangiagli20}, at which point all X-ray instruments considered in this work may be able to enclose it within their field of view.





\subsection{Current Model Limitations}
\label{sub:limitations}

\subsubsection{Systematic Uncertainties in Single Black Hole Reflection Spectroscopy}

Sections~\ref{sec:properties} and~\ref{sec:prospects} showed that evidence of SMBH binaries may be found in the combined X-ray reflection spectra from the two mini-disks. However, the properties of the predicted reflection features are almost entirely dependent on the
density structure of the disks and the details of how they are irradiated. Therefore, it is important to consider the implications of
our model assumptions and how they could impact the observational prospects of identifying SMBH binaries through X-ray spectroscopy.

A key assumption in our model setup is the presence of a static lamppost corona in both mini-disks. While observationally, the corona is consistent with being compact \citep[e.g.,][]{rm13}, the lamppost geometry is likely an idealization.  A corona that covers some fraction of the inner accretion disk will potentially produce an ionization pattern on the disk different from the one predicted by a lamppost corona, altering the emitted reflection spectrum. If the disk is overionized by an extended corona, reflection features would be weaker and harder to detect in observations. A dynamic corona will also impact the predicted reflection features in the binary spectrum. For example, an outflowing corona
\citep[e.g.,][]{beloborodov1999,dauser2013,bambic2024}, regardless of its geometry, would reduce the flux incident on the inner disk, lowering its ionization state and potentially amplifying reflection features. However, depending on the viewing angle, the outflowing corona can also beam the power-law continuum into the line-of-sight, enhancing its observed brightness and increasing the difficulty of detecting the reflection features. Therefore, an outflowing corona may either improve or decrease the chances of observing strong reflection features from a SMBH binary.

A similar give-and-take can occur with the assumed density structure of the mini-disks. The \texttt{relxilllpCp} model used here assumes a Shakura-Sunyaev density profile and though it is common in the standard thin-disk model to scale central density with the SMBH mass, we fix the central density to $10^{15}$~cm$^{-3}$ for all SMBH mini-disks (Sect.~\ref{sub:reflection_calculation}). This assumption stems from a necessity, as \texttt{relxilllpCp}'s pre-tabulated values of photoionization quantities allow for a limited range of densities between $10^{15}$ to $10^{20}$~cm$^{-3}$. 
It is worth noting that this value of the central density is appropriate for $\sim10^{6}\,M_\odot$ SMBHs, whereas central densities predicted by the standard thin-disk model for $\sim 10^{9}\,M_\odot$ SMBHs can be up to a few orders of magnitude lower. This causes a minor inconsistency in our model setup, because our calculation of the peak ionization parameters utilizes the dependence of the gas density on the SMBH mass from \citet{svensson1994}, as laid out in \citet{ballantyne2017}. As a result, the mini-disks around $\sim 10^{9}\,M_\odot$ SMBHs are characterized by higher ionizing fluxes in our models (because $F_X \propto \xi\,n_H$). Disks that are more highly ionized tend to produce ``smoother" reflection spectra with fewer features, which would make the binary signatures predicted by our model less obvious and their identification more difficult. In that sense, our model errs on the side of being more conservative. In addition to this, the X-ray reflection features in the calculated spectra are subject to poorly understood details of accretion disks and corona shared with single SMBH models. For example, winds and outflows can also sometimes complicate the analysis of an AGN reflection spectrum -- though in principle, an X-IFU observation would have the potential to be able to distinguish between outflows, winds and actual disk reflection features. Still,
continued progress in modeling the detailed radiative and gas dynamics is needed to better quantify the expected reflection signatures from single SMBHs and binaries.

\add{Both \texttt{relxill} models used to either produce the binary spectra or to subsequently fit a single SMBH model to our simulated observations are relatively simple, in that they only consider contributions from the accretion disk and the corona. Though we inject standard extinction by intervening Milky Way gas and dust, we do not include contributions from outflows, a warm corona or obscuration from gas and dust local to the SMBHs.
Real AGN X-ray spectra are notoriously diverse and complex, and idealized models (such as a `simple' \texttt{relxill} model) may not always provide a satisfactory fit, even in the absence of a binary.
Moreover, warm absorber features due to partially ionized outflows may be expected in about half of Seyfert galaxies, though their extent would be mostly limited to below $\sim 1-2$ keV.
Late-stage mergers are also expected to drive large inflows of gas and dust to the nuclear regions, increasing the likelihood and severity of obscuration compared to single AGNs \citep{hopkins2006,ricci2017,derosa2023}. 
However, as mentioned in Sect.~\ref{sub:fits}, potential signs of a binary may be present at higher energies (e.g. in the iron profile and the Compton hump) and could be detectable even in the presence of significant obscuration ($N_H\lesssim 10^{23}$~cm$^{-2}$). The expected spectral complexity of binary AGNs does suggest that variability of reflection features, such as the \fe\ line (Sect.~\ref{sub:variability}) will provide the the strongest evidence of binarity.}

Finally, Section~\ref{sec:properties} showed that the orbital blueshifting of the secondary SMBH can broaden the high-energy edge of the \fe\ profile, in some cases by 0.5~keV or more. 
However, such a shift is not unique to binary orbital motion and can in fact be seen in the spectra of single AGN. For example, a higher inclination angle enhances relativistic Doppler shifting from disk rotation, and in strongly ionized single SMBH spectra ($\log \xi \geq 2.8$), the \fe\ line is dominated by Fe\textsc{xxv} emission at 6.7 keV \citep{fabian2000,kallman2004}. Thus, an apparent shift of the blue edge in a single-epoch spectrum is not, on its own, a definitive binary signature. However, if the shift is particularly pronounced (which can be expected from higher inclination angles) or periodic (as determined from the multi-epoch spectra), conventional single-SMBH explanations may be insufficient, making the binary hypothesis more compelling.

\subsubsection{Treatment of Binary Effects}\label{sub:disc_binary}

In addition to the uncertainties pertaining to single-SMBH spectral modeling, the presence of a secondary source introduces further complexities, particularly through the addition of binary dynamics, geometric and relativistic effects. To make the problem tractable, we adopted simplifying assumptions regarding binary-specific parameters and considered only a subset of the full parameter space. Below, we discuss these choices in the context of our modeling approach.

We assume that the two mini-disks are co-planar with the binary orbital axis. This assumption is appropriate in scenarios where the angular momentum of the gas in the circumbinary disk is well defined and the binary orbit has had sufficient time to align with it through gravitational torques \citep[e.g.,][]{bogdanovic2007,miller2013}. This specific geometry minimizes the effects of cross-illumination, where emission from one corona could in principle be reprocessed by the other mini-disk. 
\add{Given our assumed coronal height of 10~\rg\ and SMBH separation of 100~\rg, the contribution of one SMBH's coronal emission to its companion's radial emissivity profile will be small in a co-planar geometry. Therefore, the impact of cross-illumination in our current setup will have a negligible effect on our results.}
For misaligned mini-disks, the incidence angle of ionizing radiation from the companion SMBH is larger, which may introduce non-axisymmetric ionization patterns in the mini-disks and, potentially, the circumbinary disk \citep{nguyen16}. As a result, the X-ray spectra in those cases may be more readily distinguishable from those of AGN powered by a single SMBH \citep[as shown by][for optical spectra]{nguyen19,nguyen20}.


We also assume that the accretion rate through the two mini-disks is steady in time. Numerical simulations have found that for $q\gtrsim 0.1$, SMBH binaries can 
have enhanced periodicity in accretion rates onto the mini-disks on approximately the orbital time scale of the binary \citep[e.g.,][]{farris2015}. Whether the time-dependent
accretion rate through the accretion streams onto the outer mini-disks propagates through to the accretion rate onto the black holes at the ISCO or is buffered by the mini-disks will depend on the spatial extent and properties of the mini-disks. To evaluate which scenario applies to the mini-disks considered in this work, we estimate the viscous time scale of the smallest mini-disk in the model and compare it to the binary orbital time. According to equation~\ref{eq:rout} and the estimates thereafter, the most compact mini-disk with $r_{\rm out} \approx 13\,r_g$ arises around the secondary SMBH in $q=0.1$ binary. The viscous time scale of this disk can be estimated as $t_{\rm visc} = -r_{\rm out}/\dot{r}_{\rm visc} \approx 16.4\,{\rm days}$, for the total mass of the binary $10^6\,M_\odot$ and accretion rate on the secondary of $\lambda_2 = 0.5$. Here $\dot{r}_{\rm visc} = -1.5 \alpha \,(h/r)^2\, v_{\rm Kep}$ is the viscous inflow rate at the radius $r_{\rm out}$ of the mini-disk, $\alpha = 0.1$ is the dimensionless viscosity parameter,  $v_{\rm Kep}$ is the circular speed of the gas in the disk, $h/r$ is the geometric aspect ratio of the standard, radiation-pressure supported thin-disk \citep{shakurasunyaev1973}. Since $t_{\rm visc} \gg t_{\rm orb}$ (see equation~\ref{eq:torbital}) for this and all other considered configurations, we conclude that we are in the regime where the accretion rate variability though the streams will be buffered by the mini-disks, due to their long viscous time scales. The assumption of steady accretion rates onto the SMBHs therefore appears to be reasonable.

We have limited our analysis to scenarios where both SMBHs are expected to be accreting through a standard geometrically-thin, optically-thick disk. However, a non-negligible portion of the parameter space shown in Fig.~\ref{fig:eddington_contour} is occupied by configurations with either a super-Eddington secondary mini-disk or an extremely sub-Eddington primary mini-disk, regimes where the thin-disk picture is invalid. 
The X-ray spectral properties of these configurations also merit investigation but are outside of the scope of this work.

The shape of the \fe\ lines and the other reflection features predicted by the \texttt{relxilllpCp} model depends not only on the ionization properties of the mini-disks, but also \add{on} the spins of both SMBHs in the binary. We have assumed near-maximal spins ($a=0.99$) for each SMBH \add{as a fiducial choice}, which is roughly consistent with observations of bright AGNs \citep[e.g.,][]{reynolds2021}.
High spins enhance the effects of relativistic blurring, and by decreasing the radius of closest stable orbit, allow emission from the deepest parts of the potential well, which is most strongly redshifted. 
However, there is some tentative evidence indicating that $a$ may be systematically lower ($a \sim 0.7$) for SMBHs with masses $\gtrsim 10^8$~M$_{\odot}$ \add{\citep[see, e.g.,][]{dotti2013}}, which would be applicable to PTA binaries. A lower SMBH spin will increase the ISCO and reduce the amount of gravitational redshift and relativistic blurring affecting the reflection spectra produced by the mini-disks.
If one or both of the SMBHs in a binary has a lower spin, then this could make the reflection features more prominent, improving the likelihood of finding evidence for the binary in the composite spectrum. We defer the exploration of a wider range of SMBH spin values to future work.

A small error is introduced in the computation of the special relativistic Doppler shift, due to the approximate way velocities are combined. In our calculation, the Doppler shift of each gas parcel in the mini-disks producing the reflection spectrum is divided into two separate components: (1) one due to the motion of the gas in the mini-disks relative to each \add{SM}BH's rest-frame, computed by \texttt{relxill}, and (2) due to the binary orbital motion, which we compute separately. By computing these Doppler shifts sequentially, we are effectively adding the bulk velocity of the mini-disks to the Keplerian motion of the gas within them in a Galilean fashion. This introduces an error of the order of a few percent to our calculation. For example, consider an emitting fluid element located in a mini-disk at a radius of 18\,\rg\ from its MBH. For a $q=1$ binary at an orbital separation of 100 \rg\ (where the orbital velocity of the binary is $0.1c$), the maximum difference in the proper velocity of the fluid between a Galilean velocity transformation and the correct special relativistic treatment through Lorentz transformations is about 1.7\%. This level of error remains within acceptable limits for the purposes of our analysis.

\section{Conclusions}
\label{sec:concl}
X-ray reflection signatures are important features in the spectra of AGNs, providing general information on the physics of accretion disks, as well as probing the space-time geometry around the central SMBH. It is therefore reasonable to expect that the reflection signal from the two SMBHs in a binary will show evidence of the presence of the binary, and could even allow measurements of the spins of SMBHs. In this paper, we developed a model to predict the reflection spectra of SMBH binaries at an orbital separation of 100\,$r_g$ with different mass ratios and Eddington accretion rates. Limiting ourselves to the regime of geometrically-thin, optically-thick mini-disks, we calculated the relativistic ionized reflection spectra from both mini-disks and produced a combined X-ray spectrum from the binary. Our main findings are as follows:

\begin{itemize}
    \item The reflection features are strongly influenced by the accretion-inversion phenomenon expected in SMBH binaries, which results in a wide range of ionization conditions in the mini-disks and the resulting composite reflection spectrum. The impact of the binary is not uniform across the composite spectrum -- in some scenarios, only the high energy reflection hump is enhanced, while in others, the impact is isolated to a mixture of soft X-ray emission lines. The SMBH binary configurations investigated here also give rise to the variable double-peaked relativistic \fe\ line profiles and a shift of the blue edge of the line amounting to about 0.5\,keV associated with their orbital motion (Section~\ref{sec:properties}).
    \item To understand whether these features are distinct enough to be used as smoking guns for the presence of SMBH binaries, we calculated mock 100\,ks observations of one of the configurations with \athena, \axis, \hexp, and \strobex\ and fit them with a single AGN model (Section~\ref{sec:prospects}). We find that, in the absence of very high-count rate data when the fit fails obviously, with a poor $\chi^2$ statistic, evidence for a binary may also be found from unusual parameter values when fitting a single-epoch spectrum. 
    
    \item For a nearby PTA binary ($10^9$\,M$_{\odot}$ at $z=0.1$) the single AGN model most obviously failed to fit the data, with strong residuals between 0.5 and 0.7\,keV and weak residuals around the \fe\ line. If this binary is instead at $z=1$, the decreased count rate masks most signs of binarity and the single AGN model becomes an acceptable fit to the mock data -- even with the large collecting area of \athena. The fits at $z=1$ were however all unable to constrain an SMBH spin in the mock data, indicating that the altered shape of the \fe\ line is preventing the fitting model from converging to a single value.

    \item For fainter SMBH binaries that are progenitors of \lisa\ sources ($10^6$\,M$_{\odot}$), spectral fitting with $100$-ks observations is only possible for a source at $z=0.1$. The single AGN model also provides a good fit to the single-epoch mock spectrum in this case but with similar challenges in constraining the SMBH spin as for the PTA binaries with $z=1$.

    \item In binaries with orbital time longer that $100$\,ks observations, multi-epoch exposures can capture spectral variability caused by the binary orbital motion. The shape of the \fe\ line is predicted to change in this sequence of spectra, and, indeed, we find that high count-rate \athena\ spectra of a PTA binary at $z=0.1$ show significant epoch-to-epoch changes in the SMBH spin when fit with a single AGN model. Such a result, combined with soft X-ray residuals that also vary with each exposure, could be compelling evidence for the presence of a SMBH binary. In the case of the \lisa\ binary configurations considered here, $100$\,ks exposure is longer that the orbital time, resulting in the averaging and loss of spectral variability signatures associated with their orbital motion.
    
\end{itemize}

Reflection spectra from accretion disks are sensitive to the assumed density structure of the disk and details of how they are illuminated by the X-ray emitting corona. In addition, there remain uncertainties in the properties of mini-disks around the two SMBHs, their relative geometries, and their interaction with the surrounding CBD. The results presented here should be considered an initial proof-of-concept study that will be expanded upon in future work. Nevertheless, it is clear that reflection signatures in the X-ray spectra of SMBH binaries will contain evidence for the presence of the binary. Deep X-ray observations of such sources may be able to unravel the components from both mini-disks, allowing insight into the evolution of SMBH binaries prior to merger.





\begin{acknowledgements}
J.M., T.B., D.R.B., T.D. acknowledge support from the NSF grant
AST-2307278 and L.B. from the NSF grant
AST-2307279. D.R.B. is also supported from NASA award
80NSSC24K0212 and NSF grant AST-2407658. T.D. acknowledges support from the DFG research unit FOR 5195 (project number 443220636, grant number WI 1860/20-1).
\end{acknowledgements}

\bibliography{refs.bib}{}
\bibliographystyle{aasjournal}

\begin{appendix}
\restartappendixnumbering 

\section{Instrument simulation files and characteristics}

\begin{deluxetable*}{cl}[h]
\tabletypesize{}
\tablewidth{0pt} 
\tablecaption{Response, ancillary and background files used to
  simulate $100$~ks observations of the SMBH binary model (e.g.,
  Fig.~\ref{fig:spectrum}). All simulations are generated using the
  \texttt{fakeit} command in \xspec. \label{tab:instrument_files}}
\tablehead{
\colhead{Instrument} & \colhead{Response files} 
} 
\startdata
\multicolumn{2}{c}{\athena} \\
\cline{1-2} 
\\[-3pt]
{} & \texttt{athena\_wfi\_rmf\_v20230523.rmf} \\
{WFI} & \texttt{NewAthena\_WFI\_13rows\_FD\_w\_filter\_OnAxis\_20240209.arf} \\
{} & \texttt{NewAthena\_WFI\_13rows\_FD\_20240223\_bkgd\_sum\_1amin\_w\_filter\_OnAxis.pha}\\[+3pt]
{} & \texttt{athena\_xifu\_3eV\_gaussian.rmf} \\
{X-IFU} & \texttt{athena\_xifu\_13\_rows\_thick\_optical\_filter.arf} \\
{} & \texttt{athena\_xifu\_nxb\_1pix.pha}\\ 
\\[-3pt]
\cline{1-2}
\multicolumn{2}{c}{\axis} \\
\cline{1-2} 
\\[-3pt]
{} & \texttt{axis\_ccd\_20221101.rmf} \\
{XMA} & \texttt{axis\_onaxis\_20230701.arf} \\
{} & \texttt{axis\_nxb\_FOV\_10Msec\_20221215.pha}\\ 
\\[-3pt]
\cline{1-2} 
\multicolumn{2}{c}{\hexp} \\
\cline{1-2} 
\\[-3pt]
{} & \texttt{HEXP\_LET\_v09.rmf} \\
{LET} & \texttt{HEXP\_LET\_v09\_PSFcor.arf} \\
{} & \texttt{HEXP\_LET\_v09\_L1\_R8arcsec.bkg} \\ [+3pt]
{} & \texttt{HEXP\_HET\_v09.rmf} \\
{HET} & \texttt{HEXP\_HET\_v09\_PSFcor\_x2.arf} \\
{} & \texttt{HEXP\_HET\_v09\_x2\_L1\_R18arcsec\_Ln25keV.bkg} \\
\\ [-3pt]
\cline{1-2}
\multicolumn{2}{c}{\strobex} \\
\cline{1-2} 
\\[-3pt]
{} & \texttt{lema\_2017-08-11.rmf} \\
{LEMA} & \texttt{lema\_60\_2023-11-16.arf} \\
{} & \texttt{lema\_nxb\_plus\_cxb.bkg} \\ [+3pt]
{} & \texttt{STROBEX\_HEMA\_300eV.rmf} \\
{HEMA300} & \texttt{STROBEX\_HEMA\_300eV.arf} \\
{} & \texttt{STROBEX\_HEMA\_300eV.bkg} \\
\\ [-3pt]
\enddata
\end{deluxetable*}

\begin{deluxetable*}{ll rrr}[h]
\tabletypesize{}
\tablewidth{0pt} 
\tablecaption{Nominal bandpass, collecting area and energy resolution of each proposed instrument included in this study. \label{tab:instrument_capabilities}}
\tablehead{
\colhead{\multirow{2}*{Mission}} & \colhead{\multirow{2}*{Instrument}} & \colhead{Nominal} & \colhead{Collecting area} & \colhead{\multirow{2}*{Energy resolution}}\\[-7pt]
\colhead{} & \colhead{} & \colhead{bandpass} & \colhead{at 6 keV} & \colhead{}
} 
\startdata
\\[-7pt]
\multirow{2}*{\athena}  & {WFI}   & 0.1-15~keV & $1,780~\textrm{cm}^2$ & 80~eV (at 1~keV), <170~eV (at 7 keV) \\
                        & {X-IFU} & 0.2-12~keV & $1,220~\textrm{cm}^2$ & 4 eV (at 7 keV) \\
\\[-7pt]
\cline{1-5}
\\[-7pt]
\axis                   & {XMA}   & 0.3-10~keV & $830~\textrm{cm}^2$ & 70~eV (at 1~keV), 150~eV (at 6~keV) \\
\\[-7pt]
\cline{1-5} 
\\[-7pt]
\multirow{2}*{\hexp}    & {LET}   & 0.2-20~keV & $530~\textrm{cm}^2$ & 80~eV (at 1~keV), 150~eV (at 7~keV) \\
                        & {HET}   & 2-80~keV & $140~\textrm{cm}^2$   & 500~eV (below 10~keV), 850~eV (at 60~keV) \\
\\[-7pt]
\cline{1-5} 
\\[-7pt]
\multirow{2}*{\strobex} & {LEMA}    & 0.2-12~keV &  $2,980~\textrm{cm}^2$ & 85~eV (at 1~keV) \\
                        & {HEMA300} & 2-30~keV   & $32,990~\textrm{cm}^2$ & 300-500~eV (at 6.4 keV) \\
\\[-7pt]
\enddata
\tablecomments{Numbers for the collecting area are sourced directly from the \texttt{.arf} files for each instrument. References for the nominal bandpass and energy resolution are taken from the following sources: \citet{meidinger2020} and \citet{cruise2025} for \athena, \citet{reynolds2023} for \axis, \citet{madsen2024} for \hexp\ and \citet{ray2024} for \strobex.}
\end{deluxetable*}
\end{appendix}



\end{document}